\documentclass{emulateapj}

\newcommand{\arcsd}[2]{$#1^{\prime\prime}\!\!.#2$}           
\newcommand{\arcmn}[2]{$#1^{\prime}\!\!.#2$}             

\newcommand{\etal}{{et al.}\/ }
\newcommand{\eg}{e.g. }
\newcommand{\ie}{i.e. }

\newcommand{\pname}{Gemini/HST Galaxy Cluster Project}
\newcommand{\rxj}{RXJ0142.0+2131}

\newcommand{\hst}{{\em HST}}
\newcommand{\rosat}{ROSAT}
\newcommand{\chandra}{{\em Chandra}}
\newcommand{\xmm}{XMM-{\em Newton}}
\newcommand{\galfit}{GALFIT}

\newcommand{\cosmology}{$H_0 = 70$ km s$^{-1}$ Mpc$^{-1}$,
$\Omega_{\mathrm{m}} = 0.3$, $\Omega_{\Lambda} = 0.7$}

\newcommand{\afe}{$[\alpha/\mathrm{Fe}]$}
\newcommand{\zform}{$z_{\mathrm{form}}$}

\newcommand{\gfil}{$g^{\prime}$}
\newcommand{\rfil}{$r^{\prime}$}
\newcommand{\ifil}{$i^{\prime}$}

\newcommand{\HdeltaA}{H$\delta_{\mathrm{A}}$}

\newcommand{\CNtwo}{CN$_{2}$}
\newcommand{\HgammaA}{H$\gamma_{\mathrm{A}}$}

\newcommand{\HdgA}{H$\delta_{\mathrm{A}} + \mathrm{H}\gamma_{\mathrm{A}}$}

\newcommand{\Hbetaem}{H$\beta_{\mathrm{G}}$}

\newcommand{\mgb}{Mg$b$}

\newcommand{\fe}{$\langle$Fe$\rangle$}

\shorttitle{The galaxy content of \rxj} 
\shortauthors{Barr et al.}

\begin{document}

\title{\rxj: I. The galaxy content of an X-ray-luminous galaxy cluster
at \lowercase{$z=0.28$}}


\author{Jordi Barr, Roger Davies}
\affil{Oxford University, Keble Road, Oxford, OX1 3RH, UK}
\email{jmb@astro.ox.ac.uk}\email{rld@astro.ox.ac.uk}

\author{Inger J\o rgensen}
\affil{Gemini Observatory, Hilo, Hawaii, USA}
\email{ijorgensen@gemini.edu}

\author{Marcel Bergmann}
\affil{NOAO Gemini Science Center, La Serena, Chile}
\email{mbergmann@noao.edu}

\and

\author{David Crampton}
\affil{Herzberg Institute of Astrophysics, Canada}
\email{david.crampton@nrc.ca}

\begin{abstract}

We present a photometric and spectroscopic study of stellar
populations in the X-ray-luminous cluster of galaxies \rxj \ at
$z=0.280$. This paper analyses the results of high signal-to-noise
spectroscopy, as well as \gfil-, \rfil-, and \ifil-band imaging, using
the Gemini Multi-Object Spectrograph on Gemini North. Of 43
spectroscopic targets, we find 30 cluster members over a range in
color. Central velocity dispersions and absorption-line strengths for
lines in the range $3700$\AA \ $ \lesssim \lambda_{\mathrm{rest}}
\lesssim 5800$\AA \ are derived for cluster members, and are compared
with a low-redshift sample of cluster galaxies, and single stellar
population (SSP) models. We use a combination of these indicators to
estimate luminosity-weighted mean ages, metallicities ([M/H]), and
$\alpha$-element abundance ratios (\afe).

\rxj \ is a relatively poor cluster and lacks galaxies with high
central velocity dispersions. Although the red sequence and the
Faber-Jackson relation are consistent with pure passive evolution of
the early-type population with a formation redshift of \zform \
$\simeq 2$, the strengths of the 4000\AA \ break and scaling relations
between metal line indices and velocity dispersion reject this model
with high significance. By inverting SSP models for the \Hbetaem,
\mgb, and \fe \ line indices, we calculate that, at a given velocity
dispersion and metallicity, galaxies in \rxj \ have
luminosity-weighted mean ages $0.14 \pm 0.07$ dex older than the
low-redshift sample. We also find that \afe \ in stellar populations
in \rxj \ is $0.14 \pm 0.03$ greater than at low redshift. All scaling
relations are consistent with these estimated offsets.

We speculate that the older luminosity-weighted mean ages and \afe \
enhancement can be brought about by a rapidly-curtailed burst of star
formation in \rxj, such as may be experienced in a cluster--cluster
merger. We note that the cluster's velocity dispersion, $1278 \pm
134$ km s$^{-1}$, is larger than expected both from its X-ray
luminosity and richness. However, the velocity distribution of
galaxies in \rxj \ is consistent with being drawn from a Gaussian
distribution and no sign of substructure is found. We conclude that
stellar populations in \rxj \ cannot evolve into stellar populations
similar to those seen in our low-redshift sample through passive
evolution. This study provides further evidence that a more complex
model, possibly involving ongoing or intermittent star formation and
galaxy mergers, is required to describe the evolution of cluster
galaxies.

\vspace{0.5cm}

\end{abstract}


\section{Introduction}

Studies of the kinematic properties of galaxies in local clusters have
revealed a number of empirical scaling relations. Primary among these
is the Fundamental Plane (FP) of elliptical (E) and lenticular (S0)
galaxies, (\eg Djorgovski \& Davis 1987; Dressler \etal 1987; J\o
rgensen \etal 1996)\nocite{djorgovski87,dressler87,jorgensen96}, which
relates surface brightness, effective radius and central velocity
dispersion in cluster members. This can be projected to give a
correlation between luminosity and velocity dispersion, the
Faber-Jackson (FJ) relation \cite{faber76}. For spiral galaxies, the
Tully-Fisher (TF) relation \cite{aaronson86} gives the relationship
between luminosity and rotational velocity. Also well-studied is the
red sequence of early-type galaxies (\eg Bower \etal 1992; Kodama \&
Arimoto 1997; Gladders \etal 1998; Smail \etal 1998; Bell \etal
2004)\nocite{bower92,kodama97,gladders98,smail98,bell04} as well as
scaling relations between velocity dispersion and the strengths of
absorption lines (\eg Bender \etal 1993; J{\o}rgensen 1997; Colless
\etal 1999)\nocite{bender93,jorgensen97,colless99}. It has been
claimed that such correlations imply that the cluster population
shares a common, often quiescent, star formation history. This must be
reconciled with observations of galaxy clusters which show complex
processes such as mergers, bursts of star formation and interactions
involving powerful active galaxies (\eg Fabian \etal 2000; Kempner
\etal 2002; Sakai \etal 2002; Owen \etal
2005)\nocite{fabian00,sakai02,kempner02,owen05}.

In order to constrain the star-formation history in clusters it is
desirable to make observations at a number of epochs, and thus
directly examine the scaling relations as a function of
redshift. There are a growing number of studies which aim to do this
(\eg van Dokkum \& Franx 1996; Ziegler \& Bender 1997; Bender \etal
1998; J\o rgensen \etal 1999; Kelson \etal 2000; van Dokkum \etal
2001; Ziegler \etal 2001; Andreon \etal 2004; Wuyts \etal
2004)\nocite{vandokkum96,ziegler97,bender98,jorgensen99,kelson00,vandokkum01,ziegler01,andreon04,wuyts04}. For
the most part, these investigations find that the amount of evolution
in a particular observable from distant clusters to a nearby reference
sample is consistent with pure passive evolution. This means that
changes can be explained entirely by a stellar population created
instantaneously at high redshift (typically \zform $>2$) and evolving
with no new star formation.

Until recently these studies typically sampled $\sim 10$ galaxies per
cluster over a narrow range in luminosity from the red sequence. This
restricts the analysis to a narrow range in galaxy mass, and so the
dependence of a particular property on mass cannot be
examined. Furthermore, observing only early-type galaxies does not
account for morphological differences between distant and nearby
clusters. It has been shown that between $z=0.5$ and $z=0$, a
significant portion of the spiral galaxies in clusters become E and S0
galaxies \cite{dressler97,vandokkum01}. This means that any study
which intends to address changes in clusters will have to avoid this
so called ``progenitor bias'' by sampling the galaxy population as a
whole rather than a potentially evolving subset.

In this work we present photometric and spectroscopic analysis of the
stellar populations in \rxj, a galaxy cluster at $z=0.28$. It is the
first of two papers on \rxj: Hubble Space Telescope (\hst) \
observations will be detailed in a future study (Barr et al., in
preparation). The present paper is itself second in a series based on
observations made as part of the \pname, targeting the stellar
populations in massive clusters of galaxies. The \pname \ sets out to
address the issues described above by observing $30 - 50$ cluster
members in 15 galaxy clusters in the $0.2 < z < 1$ interval. This is
accomplished through \hst \ imaging and deep spectroscopy from the
twin 8m Gemini telescopes, and using recent single stellar population
(SSP) models to derive luminosity-weighted mean ages, metallicities
and $\alpha$-element abundance ratios. The main science aims of the
project are outlined in J{\o}rgensen \etal (2005; hereafter
J05)\nocite{jorgensen05}. Details of the galaxy clusters observed as
part of \pname \ will be published in a future paper (J\o rgensen et
al., in preparation); the intent is to study massive clusters of
galaxies over a redshift interval approximately equal to half the age
of the universe. Clusters in the \pname \ are selected from a variety
of surveys to be representative of the population with $L_{X} \geq 2
\times 10^{44}$ erg s$^{-1}$.

We use data from three low-redshift clusters as a control
sample. These have been chosen based on the similarity of data quality
and content. Photometry and velocity dispersions for 116 galaxies in
the Coma cluster ($z = 0.023$) comes from J{\o}rgensen
(1999)\nocite{jorgensen99}. Velocity dispersions and line indices from
63 galaxies in the Perseus cluster ($z=0.018$) and 17 galaxies in
Abell 194 ($z=0.018$) are also used. The comparison galaxies lie on
the red sequence and are classified as E or S0. There are no
discernible offsets in the average measured properties of stellar
populations in these three clusters. See J05 for further details. The
full low-redshift comparison sample will be published in a future
paper (J{\o}rgensen 2005, in preparation).

In order to quantify further the changes in observables between
$z=0.28$ and $z \sim 0$, we use the models of Thomas \etal (2003a,
2004)\nocite{thomas03,thomas04}. They derive the line indices in the
Lick/IDS system for stellar populations with varying age, metallicity
([M/H]), and $\alpha$-element abundance ratio (\afe). These are
combined with the mass-to-light ratios (M/L) of Maraston
(2004)\nocite{maraston04}, the derivation of which employs the Thomas
models. For non-Lick indices we also adopt models from Vazdekis
(1997)\nocite{vazdekis97}, and Bruzual \& Charlot
(2003)\nocite{bruzual03} though these provide only for a solar \afe.
We investigate how the luminosities and values of line indices differ
for galaxies in \rxj \ compared with the low-redshift comparison
sample. We also make predictions of how these quantities would evolve
from $z = 0.28$ to $z \sim 0$ assuming pure passive-evolution with a
formation redshift of \zform $\simeq 2$. This formation redshift is
implicit when we say ``assuming passive evolution'', or ``the passive
evolution model''. Furthermore, any predicted changes in line index
measurements within this scheme are due solely to changes in the
luminosity-weighted mean ages of the stellar population. See J05 for
an in-depth description of the application of the SSP models as part
of the \pname.

Previous observations of \rxj \ are summarised in
Section~\ref{sec:rxj}. Section~\ref{sec:obs} outlines the observations
and data reduction on which this paper is based. The analysis of
derived kinematic and absorption-line quantities is the subject of
\S\ref{sec:ana}, while implications are discussed in
\S\ref{sec:dis}. Conclusions are presented in \S\ref{sec:con}. Our
adopted cosmology is \cosmology.

\section{\rxj : Background information}
\label{sec:rxj}
  
RXJ0142.0+2131 was first identified as a bright extended X-ray source
in the \rosat \ All Sky Survey \cite{voges99} and subsequently as a
massive cluster of galaxies at $z=0.280$ in both the Northern \rosat \
All-Sky Galaxy Cluster Survey \citep{bohringer00} and the \rosat \
extended Brightest Cluster Sample (Ebeling \etal
2000\nocite{ebeling00})\footnote{The redshift of RXJ0142.0+2131 in
B{\" o}hringer \etal (2000)\nocite{bohringer00} is mistakenly given as
0.0696, which is the redshift of the bright, foreground spiral. The
correct redshift is given in Ebeling \etal
(2000)\nocite{ebeling00}.}. Its X-ray luminosity is $L_{X}$ ($0.1 -
2.4$ keV) $= 6.40 \times 10^{44}$ erg s$^{-1}$ in the cluster rest
frame. There have been no \chandra \ or \xmm \ observations so no
information on the morphology of the cluster X-ray gas is available.

\section{Observational data}
\label{sec:obs}

\begin{deluxetable}{lr}

  \tabletypesize{\footnotesize}
  \tablewidth{0pt}
  \tablecolumns{2}
  \tablecaption{Instrumentation\label{tab:ins}}

  \tablehead{\colhead{} & \colhead{}}

  \startdata
    
    Telescope & Gemini North \\
    Instrument & GMOS-N \\
    CCDs & $3 \times$ EEV $2048 \times 4608$ \\
    Read-out noise \tablenotemark{a} & (3.5, 3.3, 3.0) e$^{-}$ \\
    gain \tablenotemark{a} & (2.10, 2.34, 2.30) e$^{-}$/ADU \\
    Pixel scale & \arcsd{0}{0727}/pixel \\
    Field of view & \arcmn{5}{5}$\times$\arcmn{5}{5} \\
    Imaging filters & $g'$, $r'$, $i'$ \\
    Grating & B600\_G5303 \\
    Slit width & \arcsd{0}{75} \\
    Slit length & \arcsd{4}{0} -- \arcsd{15}{0} \\
    Extraction aperture & \arcsd{0}{75} $\times$ \arcsd{1}{2} \\
    $r_{ap}$ \tablenotemark{b} & \arcsd{0}{55} \\
    Spectral resolution, $\sigma$ \tablenotemark{c} & 1.464\AA \\
    Wavelength range \tablenotemark{d} & 4000--7500\AA \\
    
  \enddata

  \tablenotetext{a}{Values for the three detectors in the array}
  \tablenotetext{b}{Equivalent circular aperture; see J\o rgensen
  \etal (1995)\nocite{jorgensen95}} 
  \tablenotetext{c}{Median of the instrumental resolutions, $\sigma$,
  measured for each slit from Gaussian fits to the sky lines. Note
  that this is equivalent to 73 km s$^{-1}$ measured at 4800\AA \ in
  the rest frame of \rxj}
  \tablenotetext{d}{The exact wavelength range varies from slit to
  slit}


\end{deluxetable}

Imaging and spectroscopy of \rxj \ were obtained with the Gemini
Multi-Object Spectrograph on Gemini North (GMOS-N) in semester 2001B
as part of GMOS-N System Verification program, GN-2001B-SV-51. See
Hook \etal (2004)\nocite{hook04} for a description of
GMOS-N. Observations were made within the period UT 2001 October 20 to
2001 November 18. Table~\ref{tab:ins} gives the instrumental setup,
Tables~\ref{tab:ima} and \ref{tab:spe} summarise the imaging and
spectroscopic observations respectively. The spectroscopic
observations were obtained as 18 individual exposures of 1800s, split
between two masks. Because some objects appear in both masks, exposure
times differ according to object, and vary from 4 to 9 hours.

\begin{deluxetable}{lccr}

  \tabletypesize{\footnotesize}
  \tablewidth{0pt}
  \tablecolumns{5}
  \tablecaption{Imaging observations\label{tab:ima}}

  \tablehead{\colhead{Filter} & \colhead{Total exposure time} &
  \colhead{Image quality} & \colhead{Sky brightness} \\ \colhead{} &
  \colhead{(s)} & \colhead{FWHM (\arcsec)} & \colhead{(mag arcsec$^{-2}$)}
  }

  \startdata

    $g'$ & $6 \times 600$ & 0.67 & 21.45 \\

    $r'$ & $8 \times 300$ & 0.52 & 20.56 \\

    $i'$ & $8 \times 300$ & 0.53 & 19.87 \\

  \enddata


\end{deluxetable}

\begin{deluxetable}{lccccc}

  \tabletypesize{\footnotesize}
  \tablewidth{0pt}
  \tablecolumns{5}
  \tablecaption{Spectroscopic observations\label{tab:spe}}

  \tablehead{\colhead{Mask ID} & \colhead{Exposure time} &
  \multicolumn{3}{c}{Image quality\tablenotemark{a}} \\ \colhead{} &
  \colhead{} & \colhead{5000\AA} & \colhead{6000\AA} &
  \colhead{7000\AA} \\ \colhead{} & \colhead{(s)} &
  \colhead{(\arcsec)} & \colhead{(\arcsec)} & \colhead{(\arcsec)}}
    
  \startdata

    GN-2001B-SV-51-2 & $18000$ & 0.72 & 0.79 & 0.70 \\
    GN-2001B-SV-51-3 & $14400$ & 0.88 & 0.88 & 0.86 \\

  \enddata

  \tablenotetext{a}{FWHM measured by fitting a Gaussian in the spatial
  direction to one of the alignment stars in the mask}


\end{deluxetable}

The basic reduction of the data was made using a combination of the
Gemini IRAF\footnote{IRAF is distributed by the National Optical
Astronomy Observatories, which are operated by the Association of
Universities for Research in Astronomy, Inc. (AURA), under cooperative
agreement with the National Science Foundation. The Gemini IRAF
package is distributed by Gemini Observatory, which is operated by
AURA.}  package and custom reduction techniques written in
IRAF. Details of these routines are given either in this paper or in
J05\nocite{jorgensen05} which contains a thorough, general description
of both the photometric and spectroscopic reduction processes applied
to GMOS-N data. In the present work we outline the data reduction and
expand only where the process differs from J05.

All data taken as part of GN-2001B-SV-51 are publicly available through
the Gemini Science Archive (http://cadcwww.hia.nrc.ca/gemini/sv/).

\subsection{Derived photometric parameters}
\label{sec:dpp}

Broad-band images in \gfil, \rfil, and \ifil \ are reduced in standard
fashion, as described in the Appendix. These images are then processed
using the object detection and photometry package SExtractor
\cite{bertin96}. We use the \ifil-band image to detect objects which
are then photometrically processed in each band individually. The
catalogs are visually inspected as a final measure to ensure that
galaxies are correctly separated from one another.

We adopt the best magnitudes ({\it mag\_best}) from SExtractor as the
total magnitude of the object. Colors are calculated from aperture
magnitudes within a diameter of \arcsd{1}{34}, which is twice the
seeing FWHM in the \gfil-band. From simulated model images of galaxies
convolved with the relevant image quality we estimate the systematic
effects on the colors due to the differences in image quality. There
is no significant effect on the \rfil$-$\ifil \ colors.  For the
\gfil$-$\rfil \ and \gfil$-$\ifil \ colors, the effect is no more than
0.02. The uncorrected colors are systematically too red. This small
effect does not significantly affect our analysis of the
data. Magnitudes and colors in \rfil \ and \ifil \ are calibrated to
rest-frame $B$ for comparison with the low-redshift sample (see
Appendix, \S~\ref{sec:crf}).

The typical uncertainties on the magnitudes and colors of galaxies in
the spectroscopic sample from photon noise alone are 0.002 mag and
0.003 mag respectively. This does not account for uncertainties
introduced by the reduction pipeline. The effect on the uncertainties
in magnitude and color due to the photometric reduction is assessed in
J05. For \ifil \ $< 21.5$ typical uncertainties are 0.035 mag in
magnitude and 0.045 mag in color, while for $21.5 <$ \ifil \ $< 22.5$
the values are 0.06 and 0.07.

We separate galaxies into ``bulge-like'' or ``disk-like'' by analysing
their surface photometry in the GMOS-N \rfil-band using
\galfit\footnote{{\it
http://zwicky.as.arizona.edu/$^{\sim}$cyp/work/galfit/galfit.html}}
\cite{peng02}. The \galfit \ program fits various types of
two-dimensional luminosity profiles to data. \galfit \ is used \ to
fit each galaxy in the spectroscopic sample as a S\'ersic profile
\cite{sersic68} and find the best-fitting power-law index, $n$. We use
this determination in a relative sense to distinguish bulge-like from
disk-like and arbitrarily define anything with $n \geq 2$ as a
bulge-like galaxy.

\subsection{Spectroscopic data -- sample selection}
\label{sec:sss}

The spectroscopic sample was selected based on the photometry.  Stars
and galaxies were separated using the SExtractor classification
parameter {\it class\_star} derived from the image in the $i'$-filter.
At the time of the sample selection for RXJ0142.0+2131 we were using a
threshold of 0.90, i.e., objects with {\it class\_star} $<0.90$ in the
$i'$-image are considered galaxies.  For clusters observed later in
the project, we used 0.80. The effect of this is discussed below.

Selection categories were set based on the total magnitude in $r'$
and the colors.  We used a color selection that includes all likely
cluster members. The categories were set as follows.

\begin{itemize}

  \item 1: $r' \le 19.4 ~ \wedge ~ 1.8 \le (g'-i')\le 2.4 ~ \wedge ~
  (r'-i')\le 0.7$

  \item 2: $19.4 < r' \le 21.2 ~ \wedge ~ 1.8 \le (g'-i')\le 2.4 \\
         \wedge ~ (r'-i')\le 0.7$

  \item 3: $19.4 < r' \le 21.2 ~ \wedge ~ 1.0 \le (g'-i') < 1.8$

  \item 4: $\left( r'\le 21.2 ~ \wedge ~ (g'-i') < 1.0 \right) \\ \vee
  ~ \left( r'\le 21.2 ~ \wedge ~ ( (g'-i') > 2.4) ~ \vee ~ (r'-i')>0.7
  ) \right) \\ \vee ~ \left( 21.2< r' \le 21.6 ~ \wedge ~ 1.0 \le
  (g'-i') \le 2.4 \right)$

\end{itemize}

Figure \ref{fig:cmd} summarises the photometry for the field as
color-magnitude diagrams and color-color diagrams. The spectroscopic
sample is marked, with cluster members as solid green boxes. The
selection categories are visualized on Figure
\ref{fig:cmd}b. Positions and photometry for the spectroscopic sample
are given in Table~\ref{tab:pho}.

\begin{figure*}

  \begin{center}

    \plotone{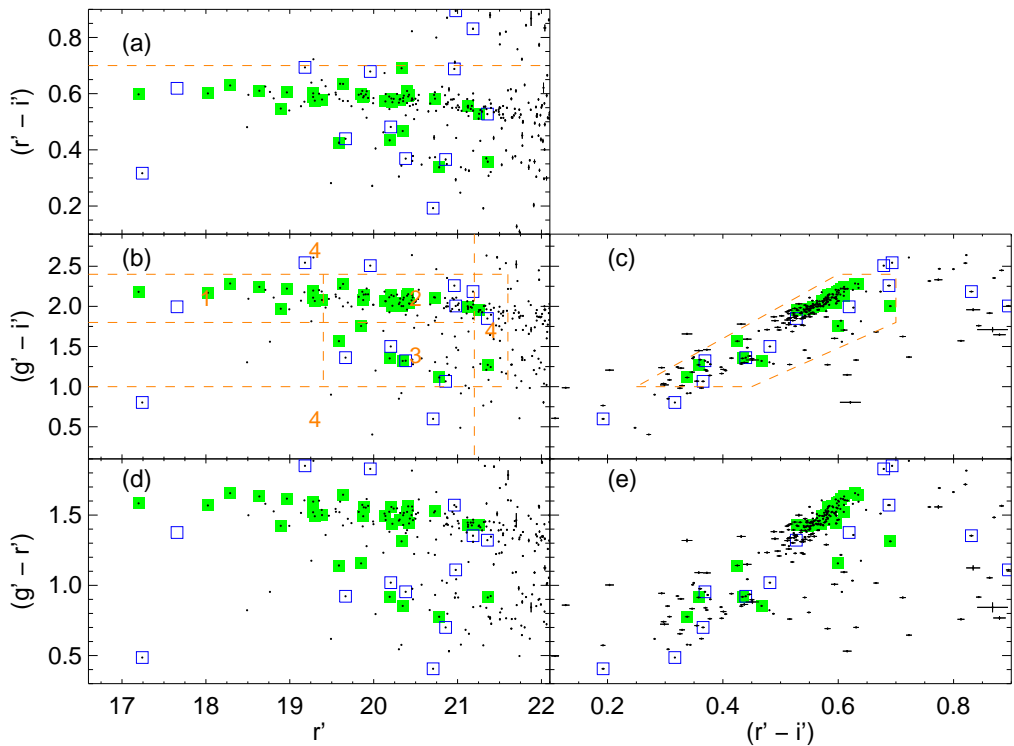}

    \figcaption{RXJ0142.0+2131: Color-magnitude and color-color
    diagrams. Objects with $class\_star \leq 0.80$ and \rfil \ $\leq
    22.1$ are marked with points and error bars. Boxes indicate those
    objects in the spectroscopic sample. Filled boxes are confirmed
    cluster members. The dashed lines show the magnitude- and
    color-based selection criteria and categories used in assigning
    spectroscopic targets are shown in the $($\gfil$-$\ifil$)$ vs.
    \rfil \ plot. The open box without a corresponding point is the M
    star whose $class\_star$ is $0.88$ (see
    \S\ref{sec:sss}).\label{fig:cmd}}

  \end{center}

\end{figure*}

All objects included in categories 1, 2, and 3 were also required to
meet the condition \\

\begin{displaymath}
  (r'-i')\le 0.7 ~ \wedge ~ 1.0 \le (g'-i') \le 2.4 ~ \wedge ~
\end{displaymath}

\begin{displaymath}
  (g'-i) \le 4 \cdot (r'-i') ~ \wedge ~ (g'-i) \ge 3.2 \cdot (r'-i') - 0.44
\end{displaymath}

This condition corresponds to the area outlined on Figure
\ref{fig:cmd}c.

Category 1 and 2 objects are the most important to include in the
spectroscopic sample. Roughly the same number of galaxies from each of
these categories were included in the final spectroscopic
sample. Category 3 objects, which are likely to include blue cluster
members, are included whenever no category 1 or 2 object is
available. Due to the distribution of the category 1, 2 and 3 objects
in the field, not all of the available space on the masks could be
filled.  We therefore included category 4 objects in order to fill
both masks.  The very blue category 4 objects are expected to be
foreground galaxies, while the very red category 4 objects may be
background galaxies.  The faint category 4 objects are expected to
include cluster members.

The star-galaxy classification parameter {\it class\_star} in the
$i'$-filter is 0.01-0.04 for all the objects in the spectroscopic
sample, except for one object which has {\it class\_star} $=0.88$ and
turned out to be an M star. If we had used the criteria {\it
class\_star} $<0.80$ as done for other clusters in the \pname, this
star would not have been included in the sample.  However, its
inclusion has not significantly changed the sample selection, since
only one other possible cluster member would have been observed
instead.

The M star is the only category 1 non-member. There are two faint
category 2 non-members.  Four of the nine category 3 galaxies observed
were non-members, while all the very blue and very red category 4
galaxies were non-members. Of the three faint category 4 galaxies
observed, two are members of RXJ0142.0+2131. Figure~\ref{fig:loc}
shows the \rfil-band image with the spectroscopic sample marked.

\begin{figure*}

  \begin{center}

    \plotone{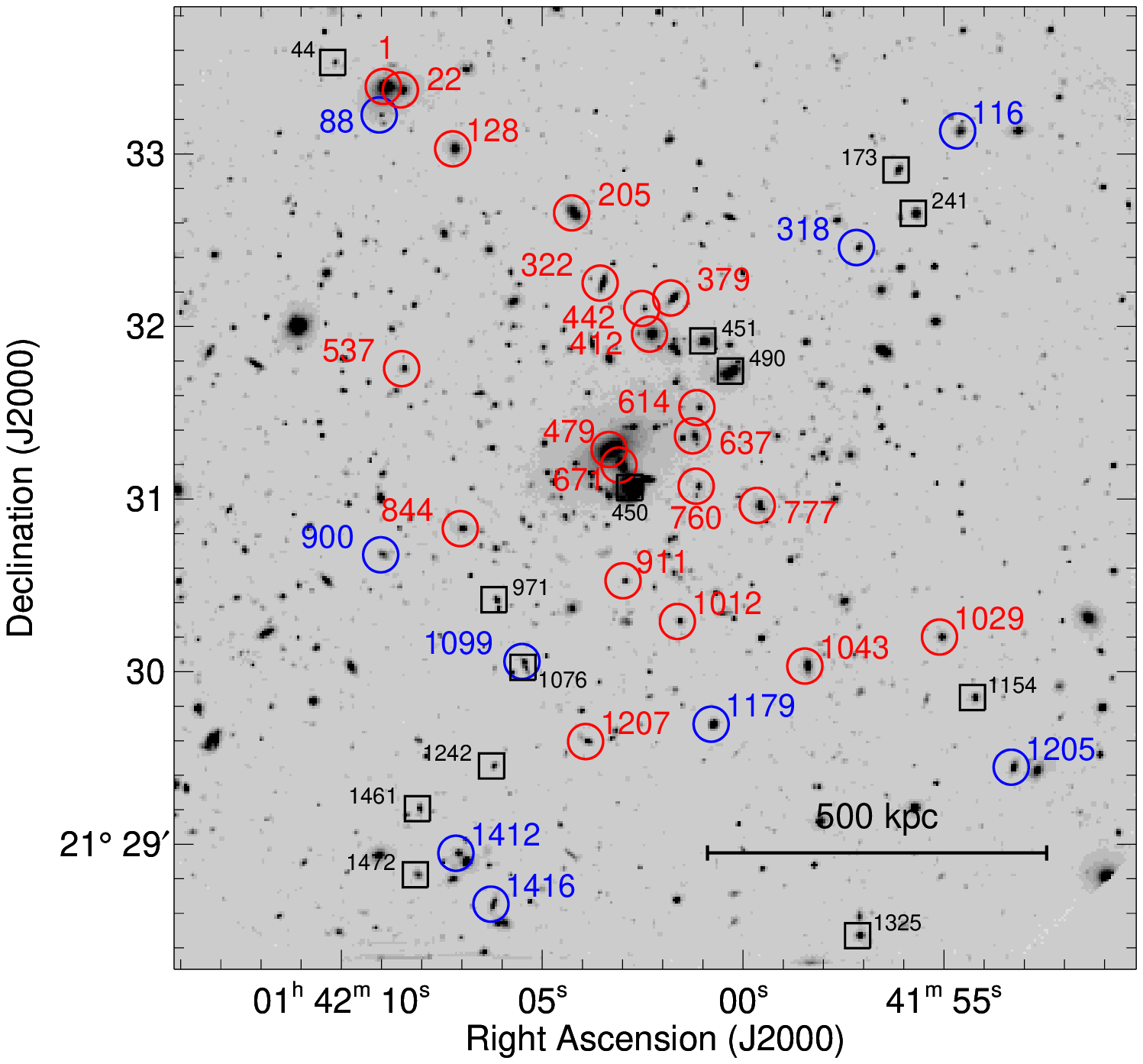}

    \figcaption{RXJ0142.0+2131 \rfil-band image with the spectroscopic
    sample marked. The field size is \arcmn{5}{5}$\times$\arcmn{5}{5};
    North is up, East is left. Confirmed cluster members are marked
    with a circle, non-members with a square. Cluster galaxies are
    color coded according to morphology; red: bulge-like; blue:
    disk-like.\label{fig:loc}}

  \end{center}

\end{figure*}

\subsection{Derived spectroscopic parameters}

Flux-calibrated one- and two-dimensional spectra are produced for each
object as described in the Appendix (\S~\ref{sec:sdr}), which follows
the method in J05.

Redshifts, velocity dispersions and line indices are derived from the
extracted 1D spectra. For galaxies with emission lines, the initial
redshift estimates are made using these. Otherwise, the spectrum is
cross-correlated with that of the K0III star HD172401, using the IRAF
tasks {\tt fxcor} and {\tt xcsao}.

With the redshift determined to $\pm 200$ km s$^{-1}$, more accurate
kinematic parameters are determined using spectral-fitting software
written by Karl Gebhardt (Gebhardt \etal 2000; 2003; see also Saha \&
Williams 1994; Merritt
1997)\nocite{gebhardt00,gebhardt03,saha94,merritt97}; see J05 for a
description of how this program is adapted to deal with
intermediate-to-high redshift spectra. The program simultaneously fits
the kinematics and optimises the template mix in pixel space using a
maximum penalised likelihood (MPL) method. The templates are
constructed from three stars observed with GMOS-N to prevent
systematic errors being introduced by template mismatch. The
observations of the template stars are detailed in J05.

For each galaxy, the spectrum is normalised (using a 27-piece cubic
spline, rejecting points $\pm 3\sigma$ from the fit), shifted to the
rest frame and cut to the wavelength covered by both galaxy and
template stars. Emission lines and sky residuals are masked. In most
cases the fits are from $3750 - 5400$\AA \ in the rest frame with
exceptions being where the S/N is too low at the blue end to be of any
use. The best fit is then determined using the MPL
technique. Uncertainties are estimated from Monte Carlo
simulations. Values of velocity dispersion, refined redshift and
template fractions are returned. The latter information can be used to
give an estimate of the spectral classification for the
galaxy. Velocity dispersions calculated from the template fitting are
corrected to an aperture of \arcsd{3}{4} at the redshift of the Coma
cluster according to the prescription of J\o rgensen \etal
(1995)\nocite{jorgensen95}. The results of the template fitting are
shown in Table~\ref{tab:kin}.

We derive the absorption line indices CN$_2$, G4300, C4668, \mgb,
Fe5270 and Fe5335 (Combined as \fe\footnote{\fe $ = 0.5
(\mathrm{Fe5270} + \mathrm{Fe5335} )$}) from Worthey \etal
(1994)\nocite{worthey94} on the Lick/IDS system. We also derive
\Hbetaem \ \cite{gonzales93,jorgensen97}, the higher-order Balmer
lines \HgammaA \ and \HdeltaA \ \cite{worthey97}, the D4000 index
\cite{bruzual83,gorgas99} and the blue indices CN3883 and CaHK
\cite{davidge94}. Line indices are determined from spectra convolved
to the instrumental resolution of the Lick/IDS system. They are then
corrected for velocity dispersion using the technique described in
Davies \etal (1993)\nocite{davies93}; see also J05. The zero-velocity
indices are corrected to an aperture of \arcsd{3}{4} at the Coma
cluster after J\o rgensen (1997)\nocite{jorgensen97}. See J05 for the
adopted aperture corrections. No correction for spectral-shape
differences between Lick/IDS and our spectra is made. Previous studies
have shown these to be small, with uncertainties almost as large as
the offsets themselves (\eg J{\o}rgensen
1997)\nocite{jorgensen97}. The line indices for the spectroscopic
sample are listed in Table~\ref{tab:lin}. Note that we do not derive
Fe4383 as residuals from the strong sky line at 5577\AA \ fall within
the line band for cluster members. This sky line also affects some of
the \HgammaA \ measurements. In these cases (IDs 322, 671, 1179, 1412)
\HdgA \ was calculated using a linear relation between \HdeltaA \ and
\HdgA \ derived from the other cluster members. There are also four
cluster members which have accurate determinations of Fe5270 (IDs 116,
128, 1029, 1076), but for which Fe5335 either lies beyond the of
spectral range or has a large error. For these objects we calculate
\fe \ by assuming a linear relation between Fe5270 and \fe \
calibrated using those cluster members with both Fe5270 and Fe5335.

Of the 43 objects targeted for spectroscopic observations, 30 are
found to be cluster members (see \S\ref{sec:cvd} and
Table~\ref{tab:kin}), 12 are galaxies at other redshifts and one is an
M star. Extracted 1D spectra of the galaxies are reproduced in the
appendix, cluster members are shown in Figure~\ref{fig:sp1} and
non-members in Figure~\ref{fig:sp2}. GMOS-N color images of each
galaxy in the form of postage stamps are shown in Figure~\ref{fig:crm}
for cluster members and Figure~\ref{fig:crn} for non-members.

\section{Results \& Analysis}
\label{sec:ana}

In this section, redshift information is used initially to determine
cluster membership. We then examine the properties of the cluster as a
whole, \eg cluster velocity dispersion and richness, and search for
substructure, before analysing diagnostics of the stellar populations
of individual galaxies. When examining stellar populations we focus on
the color-magnitude diagram, scaling relations involving the velocity
dispersion, and absorption-line indices.

\subsection{The cluster properties of \rxj}
\label{sec:cvd}

In order to determine membership of \rxj , we first exclude galaxies
$\pm 3000$ km s$^{-1}$ from the published redshift of 0.280. We then
iteratively determine $z_{\mathrm{cluster}}$ and
$\sigma_{\mathrm{cluster}}$ using the biweight distribution method
\cite{beers90}. A total of 30 galaxies are classified as cluster
members with a mean redshift of $0.2796 \pm 0.0008$ and velocity
dispersion in the rest frame $1278 \pm 134$ km s$^{-1}$. As a
comparison, we also determine $\sigma_{\mathrm{cluster}}$ using the
method of Danese \etal (1980)\nocite{danese80}. This yields a value of
$1212^{+207}_{-143}$ km s$^{-1}$, which agrees with the biweight
distribution estimate. The X-ray luminosity of \rxj \ is about 1.7
times that of Coma, which has a line-of-sight velocity dispersion of
$1010^{+51}_{-44}$ km s$^{-1}$ \cite{zabludoff90}. If we assume that
Coma lies on the $L_{X} - \sigma$ relation ($L_X \propto
\sigma^{4.4}$) of Mahdavi \& Geller (2001)\nocite{mahdavi01}, we find
that $L_{X}$ in \rxj \ is fainter than predicted by 0.22 dex in $\log
L_X$. The scatter in the Mahdavi \& Geller relation is 0.18 dex in
$\log L_X$. It would be unwise to draw firm conclusions on X-ray
luminosity from observations made with \rosat. \chandra \ and \xmm \
observations of previously classified \rosat \ clusters of galaxies
have shown \rosat \ X-ray flux values to be too high because of the
effect of unresolved AGNs (\eg Donahue \etal
2003)\nocite{donahue03}. We are therefore required to view the \rosat
\ flux measurement as an upper limit.

Figure~\ref{fig:loc} shows the spatial distribution of
spectroscopically confirmed cluster members in \rxj. Galaxies are
coded according to whether their luminosity profiles in the \rfil-band
are bulge-like or disk-like as described in
\S\ref{sec:dpp}. Bulge-like and disk-like galaxies are well separated
spatially, as we might presume from a morphology-density relation (\eg
Dressler 1980)\nocite{dressler80}. The brightest and second brightest
cluster members (ID 479 and ID 1) are separated by $\sim 650$ kpc, and
the brightest cluster galaxy (BCG) is displaced by 1000 km s$^{-1}$
from the systemic cluster redshift, which might suggest that \rxj \ is
not fully virialized. However, the spatial distribution of
red-sequence galaxies as a whole shows no obvious substructure
projected on the sky.

\begin{figure}

  \begin{center}

    \epsscale{1.2}
    \plotone{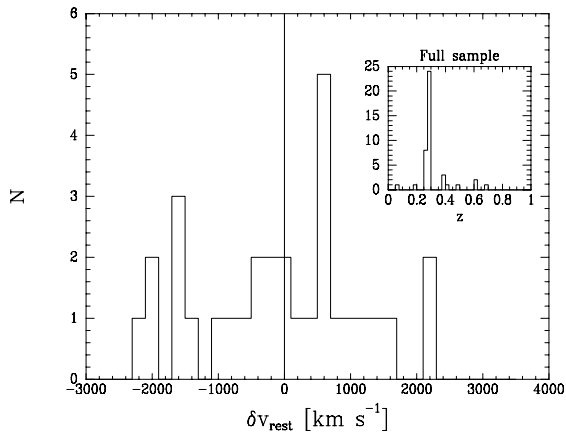}

    \figcaption{Rest-frame velocity distribution for cluster members. The
    inset shows the redshift distribution of the spectroscopic
    sample.\label{fig:zhi}}

  \end{center}

\end{figure}

The velocity distribution of cluster members is shown in
Figure~\ref{fig:zhi}. A one-sided Kolmogorov-Smirnoff test shows
the probability that the data are drawn from a Gaussian distribution is
greater than $75\%$. We can also search for substructure using the
method of Dressler \& Shectman (1988)\nocite{dressler88} which
identifies deviations in the velocities of galaxies and their
projected neighbours from the cluster as a whole. The statistic
returned ($\Delta$) is $\chi^{2}$-like in that a Gaussian distribution
returns a value of $\Delta$ which is of order of the number of degrees
of freedom. For \rxj, $\Delta = 28.9$. In order to quantify more
properly the significance of this statistic we perform a Monte Carlo
analysis on 1000 alternative realisations of the data. In each case
the velocities are shuffled randomly and reassigned and the statistic
is recalculated. We find that a higher value of $\Delta$ is found in
$>40\%$ of the artificial realisations implying that substructure is
not significant in \rxj.

In order to estimate the richness of \rxj, we determine the
galaxy-cluster spatial cross-correlation amplitude ($B_{gc}$) of Yee
\& L{\'o}pez-Cruz (1999)\nocite{yee99}. $B_{gc}$ quantifies the excess
number of objects within 0.5 Mpc of the BCG, and attempts to introduce
some redshift independence by normalising by an integrated luminosity
function. See Seldner \& Peebles (1978)\nocite{seldner78} and Longair
\& Seldner (1979)\nocite{longair79} for a full derivation of this
quantity. We employ luminosity functions from Wold \etal
(2000)\nocite{wold00}. In the absence of control observations of blank
fields, we use the area outside 0.5 Mpc of the BCG to determine a
``field density''. This means we may overestimate the background where
cluster members extend further than 0.5 Mpc from the BCG. However, we
note that photometric data from the Coma cluster (J{\o}rgensen, in
preparation) give $B_{gc}$ (Coma) $= 976 \pm 109 \: h_{50}^{-1.8}$
Mpc$^{-1.8}$, which is in agreement with the value of $1242 \pm 282 \:
h_{50}^{-1.8}$ Mpc$^{-1.8}$ published in Yee \& L{\'o}pez-Cruz
(1999)\nocite{yee99}. We find $B_{gc} (\mathrm{\rxj}) =637 \pm 132 \:
h_{50}^{-1.8}$ Mpc$^{-1.8}$.

Yee \& Ellingson (2003)\nocite{yee03} give the relationship between
$B_{gc}$ and cluster velocity dispersion for galaxy clusters at $0.2 <
z < 0.5$. They find that the two quantities are well correlated over
$500 < B_{gc} < 2000$. Our determinations of velocity dispersion and
$B_{gc}$ for \rxj \ indicate that it is an outlier, with velocity
dispersion around twice the expected value. We must be cautious with
this result as we know that the estimation of the surface density of
background galaxies may cause $B_{gc}$ to be too low. However, to make
the $B_{gc} (\mathrm{\rxj})$ consistent with the Yee \& Ellingson
prediction of $B_{gc} \simeq 2250$ we would have to reduce our
estimated background surface density by a factor of 20. We consider
such an overestimate highly unlikely as it implies a galaxy surface
density of only $\sim 500$ deg$^{-2}$ down to a magnitude of \rfil \
$=23$, a factor of $\sim 50$ lower than that found by deep wide-field
surveys (see \eg Postman \etal 1998; Wilson
2003)\nocite{postman98,wilson03}.

We also calculate the fraction of blue galaxies ($f_{B}$) in \rxj \
following Butcher \& Oemler (1984)\nocite{butcher84}. In order to make
a fair comparison, we convert our \rfil \ and \gfil$-$\rfil \
quantities into $V$ and $B-V$ via the relations of Smith \etal
(2002)\nocite{smith02}. Butcher \& Oemler define $f_{B}$ as the
fraction of galaxies with $M_{V} < -20$ and $B-V$ 0.2 mag bluer than
the red sequence. We note that there are two aspects of the
calculation that are difficult to reproduce. Firstly, $f_{B}$ is
calculated in the radius within which $30\%$ of the cluster galaxies
are contained ($R_{30}$). We find it difficult to determine this
quantity for \rxj, because the cluster is not strongly concentrated,
and our field-of-view is too small to permit an estimate of the
cluster's extent. We therefore take $R_{30}$ to be equal to the median
$R_{30}$ found by Butcher \& Oemler for clusters with $0.2 < z <
0.4$. This gives us $R_{30} (\mathrm{\rxj})= $ \arcmn{2}{4}, but we
note that the value of $f_{B}$ remains consistent if $R_{30}$ is
changed by $\pm$ \arcmn{0}{5}. More important is the correction
applied to account for foreground and background galaxies. We make
this adjustment, in the same way as for $B_{gc}$, by assuming that the
galaxies at projected distances greater that 0.5 Mpc from the BCG can
be taken as representative of the field. We use these galaxies to
determine the expected number of galaxies within $R_{30}$, centred on
the BCG, with $M_{V} < -20$ and $B-V$ 0.2 mag bluer than the red
sequence .

We find that $f_{B} (\mathrm{\rxj})= 0.22 \pm 0.06$. This is
consistent with the $f_{B} - z$ measurements of Butcher \& Oemler
(1984)\nocite{butcher84} and Fairley \etal (2002)\nocite{fairley02}
for clusters at $z \simeq 0.28$. How this figure is affected by
contamination of the background by cluster galaxies depends on the
ratio of blue and red galaxies at $>0.5$ Mpc from the BCG, and their
relative numbers when compared with the background. It is quite
difficult to gauge this effect. However, we note that the blue
fraction of these ``background'' galaxies is $0.45 \pm 0.08$,
consistent with the Butcher \& Oemler determination of the field.

The analyses of cluster velocity dispersion, X-ray luminosity,
richness, and blue fraction employ no spectroscopic information other
than redshift. As such they form only a prelude to the real meat of
this paper. Nevertheless, they suggest that \rxj \ has a higher
$\sigma_{\mathrm{cluster}}$ than both its X-ray luminosity and optical
richness would suggest, and its blue fraction is as expected from its
redshift. These phenomena are examined further in the context of
stellar populations in \S\ref{sec:dis}.

\subsection{The color-magnitude diagram}
\label{sec:cmd}

We compare the slope and zero-points of the red sequence of our GMOS-N
spectroscopic sample with the predictions of Kodama \& Arimoto (1997)
for a cluster of galaxies at $z = 0.28$. Kodama \& Arimoto conclude
that the presence of a cluster red sequence at all redshifts can be
explained if ellipticals have a common formation epoch, \zform $\simeq
2$, and evolve passively. The slope of the red sequence is caused by
fainter galaxies having lower metallicities; the timescale for the
loss of metals via a galactic wind is shorter for less massive
galaxies. Their [M/H] values range from 0.15 to $-0.37$ over $-22 <
M_{B} < - 18$ for a 15 Gyr old galaxy. More usefully, Kodama \&
Arimoto give the evolution of elliptical galaxies on the $(g-r)$ vs.
$r$ color-magnitude diagram (CMD) with redshift. We compare their red
sequence at $z=0.28$ with our data.

The $($\gfil$-$\rfil$)$ vs. \rfil \ color-magnitude diagram for \rxj
\ is shown in Figure~\ref{fig:grd}. The slope and intercept of the red
sequence are calculated by making a least-squares fit to the cluster
members with $($\gfil$-$\rfil$) > 1$, iteratively rejecting points
that deviate from the fit by more than $3\sigma$. The predicted red
sequence of a cluster of galaxies at $z = 0.28$ \cite{kodama97} is the
dotted line. The model agrees with the observed red sequence,
indicating that the colors of bulge-like galaxies in \rxj \ are
consistent with a passively-evolving population formed at \zform
$\simeq 2$.

\begin{figure}

  \begin{center}

    \epsscale{1}
    \plotone{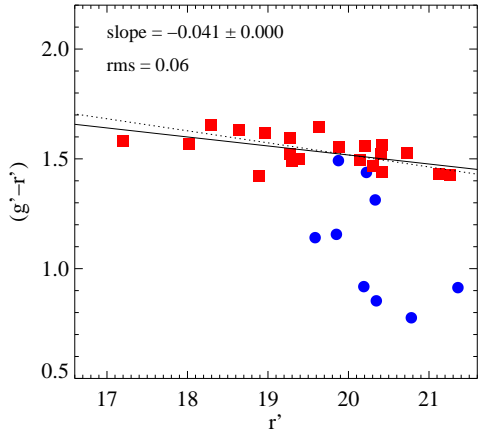}

    \figcaption{Color-magnitude diagram for spectroscopically-confirmed
    members of \rxj. Squares represent bulge-like galaxies, circles
    are the disk-like members (see the text for a description of how
    these are defined). The solid line is an iterative fit to
    red-sequence galaxies (see text), and the dotted line is the
    predicted $($\gfil$-$\rfil$)$ vs. \rfil \ red sequence at $z =
    0.28$ for a passively-evolving elliptical population formed at
    \zform $\simeq 2$ from Kodama \& Arimoto (1997).\label{fig:grd}}

  \end{center}

\end{figure}\nocite{kodama97}

Figure~\ref{fig:grd} shows cluster members divided according to
morphology. The bulge-like and disk-like galaxies are well separated;
only two of the latter appear on the red sequence. It is important to
note that the morphological determination is independent of color --
we do not assume that disk-like galaxies belong in the blue
population, or vice-versa. As noted in the previous section, the two
populations are also separated spatially.

\subsection{Correlations with velocity dispersion}
\label{sec:vds}

We now investigate how the luminosities and absorption line strengths
of galaxies in \rxj \ vary with the central velocity dispersion, and
in particular if these relationships are the same as those in the
low-redshift sample. The indices' dependence on age, metallicity and \afe \
are calculated using the SSP models of Thomas \etal (2003a,
2004)\nocite{thomas03,thomas04}, Maraston (2004)\nocite{maraston04},
Vazdekis (1997)\nocite{vazdekis97}, and Bruzual \& Charlot
(2003)\nocite{bruzual03}. The models provide H$\beta$ which is
converted to \Hbetaem \ using the transformation from J{\o}rgensen
(1997)\nocite{jorgensen97}. See J05 for a full description of the
derivation of these scaling relations. Table~\ref{tab:mod} outlines
the predicted behavior of the indices used in this paper with age,
metallicity and \afe.

\begin{deluxetable*}{llrl}

  \tablecaption{Predictions from single stellar population models
  \label{tab:mod} }

  \tablewidth{0pt}

  \tablecolumns{4}

  \tablehead{
    \multicolumn{2}{l}{Relation} & \colhead{rms} & \colhead{Reference} }

  \startdata

    $\rm \log M/L_B $ &$= \;\;\:0.935 \log {\rm age} + 0.337 {\rm
    [M/H]} - 0.053 $ & 0.022 & Maraston 2004 \\

    $\log {\rm H\beta _G} $ &$= -0.221 \log {\rm age} - 0.114 {\rm
    [M/H]} + 0.055 {\rm [\alpha/Fe]} + 0.500$ & 0.010 & Thomas et
    al. \\

    $\rm (H\delta _A + H\gamma _A)' $ &$= -0.115 \log {\rm age} -0.095
    {\rm [M/H]} +0.095 {\rm [\alpha/Fe]} + 0.009 $ & 0.008 & Thomas et
    al.\tablenotemark{a} \\

    $\rm D4000 $ &$= \;\;\:0.730 \log {\rm age} + 0.711 {\rm [M/H]} +
    1.827$ & 0.052 & Vazdekis-2000 \\

    $\log {\rm Mg}b $ &$= \;\;\:0.173 \log {\rm age} + 0.309 {\rm
    [M/H]} + 0.210 {\rm [\alpha/Fe]} + 0.354$ & 0.019 & Thomas et
    al. \\

    log \fe  &$= \;\;\:0.113 \log {\rm age} + 0.253 {\rm
    [M/H]} - 0.278 {\rm [\alpha/Fe]} + 0.343$ & 0.007 & Thomas et
    al. \\

    $\log {\rm C4668} $ &$= \;\;\:0.145 \log {\rm age} + 0.581 {\rm
    [M/H]} + 0.023 {\rm [\alpha/Fe]} + 0.529$ & 0.037 & Thomas et
    al. \\

    ${\rm CN_2} $ &$= \;\;\:0.121 \log {\rm age} + 0.196 {\rm [M/H]} +
    0.066 {\rm [\alpha/Fe]} - 0.043$ & 0.025 & Thomas et al. \\

    $\log {\rm CaHK}$ & $=\;\;\:0.073 \log {\rm age} + 0.061 {\rm
    [M/H]} + 1.291$ & 0.010 & Bruzual \& Charlot 2003 \\

    ${\rm CN3883} $ &$= \;\;\:0.173 \log {\rm age} + 0.142 {\rm [M/H]}
    + 0.086$ & 0.012 & Bruzual \& Charlot 2003  \\

    $\log {\rm G4300} $ &$= \;\;\:0.162 \log {\rm age} + 0.163 {\rm
    [M/H]} + 0.114 {\rm [\alpha/Fe]} + 0.552$ & 0.029 & Thomas et
    al. \\

  \enddata

   \tablenotetext{a}{ $({\rm H\delta _A + H\gamma _A})' \equiv
     -2.5~\log \left ( 1.-({\rm H\delta _A + H\gamma _A})/(43.75+38.75)
     \right )$, cf.\ Kuntschner (2000).  The rms for the relation
     translates to an rms of $\rm H\delta _A + H\gamma _A$ of $\approx
     0.65$ for the typical values of $\rm H\delta _A + H\gamma _A$.
    }

  \tablecomments{(1) Relation established from published model
  values. $[M/H]\equiv \log Z/Z_\sun$ is the total metallicity
  relative to solar. \afe \ is the abundance of the $\alpha$-elements
  relative to iron, and relative to the solar abundance ratio. The age
  is in Gyr. M/L is the stellar mass-to-light ratio in solar
  units. (2) Scatter of the model values relative to the relation. (3)
  Reference for the model values.}

\end{deluxetable*}\nocite{kuntschner00}

The scaling relations for \rxj \ and the comparison sample are shown
in Table~\ref{tab:sca} and Figures~\ref{fig:avs} and
\ref{fig:1vs}. Note that the values of $M_{B}$ come from the Coma
cluster, whereas values of line indices, with the exception of
\Hbetaem, \mgb, and \fe, are those of Perseus and Abell 194. The
differing spectral range of the individual spectra mean that not all
indices can be derived for all objects. For this reason the number of
points in Figures~\ref{fig:avs} and \ref{fig:1vs} differs from panel
to panel.

\begin{deluxetable*}{lllrllrlllll}

  \tablecaption{Scaling relations\label{tab:sca}}

  \tablewidth{0pt}
  \tabletypesize{\footnotesize}
  \tablecolumns{12}

  \tablehead{\multicolumn{2}{c}{Relation} &
  \multicolumn{3}{c}{Low-redshift sample} & \multicolumn{3}{c}{\rxj} &
  \colhead{$\Delta \gamma_{i}$} & \colhead{$\sigma_{\Delta \gamma \:
  i}$} & \colhead{$\sigma_{\mathrm{sys \: i}}$} & \colhead{PE$_{i}$}
  \\ \colhead{} &\colhead{} & \colhead{$\gamma_{i}$} &
  \colhead{N$_{\mathrm{gal}}$} & \colhead{rms} &
  \colhead{$\gamma_{i}$} & \colhead{N$_{\mathrm{gal}}$} &
  \colhead{rms} &\colhead{} & \colhead{} & \colhead{} & \colhead{} \\
  \multicolumn{2}{c}{(1)} & \colhead{(2)} & \colhead{(3)} &
  \colhead{(4)} & \colhead{(5)} & \colhead{(6)} & \colhead{(7)} &
  \colhead{(8)} & \colhead{(9)} & \colhead{(10)} & \colhead{(11)}}

  \startdata

    M$_{\mathrm{B}}$ & $ = (-8.02 \pm 1.08) \; \mathrm{log} \sigma +
    \gamma_{i}$ & $-2.29$ & 116 & 0.81 & $-2.94$ & 21 & 1.06 & $-0.65$
    & $0.24$ & $0.21$ & $-0.35$ \\

    log H$\beta_{\mathrm{G}}$ & $= (-0.25 \pm 0.05) \; \mathrm{log}
    \sigma + \gamma_{i}$ & $\;\;\:0.870$ & 160 & 0.086 & $\;\;\:0.850$ &
    21 & 0.093 & $-0.020$ & $0.021$ & $0.007$ & $\;\;\:0.033$ \\

    H$\delta_{\mathrm{A}} + \mathrm{H}\gamma_{\mathrm{A}}$ & $= (-9.1
    \pm 1.0) \log \sigma + \gamma_{i}$ \tablenotemark{a} & $\;13.16$ &
    65 & 1.53 & $\;13.24$ & 21 & 3.74 & $\;\;\:0.08\;\:$ & $0.84$ &
    $0.24$ & $\;\;\:1.30$ \\

    D4000 & $= \gamma_{i}$ & $\;\;\:2.10$ & 65 & 0.16 & $\;\;\:2.28$ &
    15 & 0.26 & $\;\;\:0.18\;\:$ & $0.07$ & \nodata & $-0.11$ \\

    log \mgb \ & $= (0.31 \pm 0.02) \; \mathrm{log} \sigma +
    \gamma_{i}$ & $-0.063$ & 144 & 0.059 & $-0.022$ & 21 & 0.098 &
    $\;\;\:0.041$ & $0.022$ & $0.008$ & $-0.026$ \\

    log \fe & $= (0.16 \pm 0.03) \; \mathrm{log} \sigma + \gamma_{i}$
    & $\;\;\:0.103$ & 144 & 0.053 & $\;\;\:0.096$ & 20 & 0.049 &
    $-0.007$ & $0.012$ & $0.004$ & $-0.017$ \\
   
    log C4668 & $= (0.33 \pm 0.08) \; \mathrm{log} \sigma +
    \gamma_{i}$ & $\;\;\:0.107$ & 65 & 0.058 & $\;\;\:0.117$ & 20 &
    0.150 & $\;\;\:0.010$ & $0.034$ & $0.009$ & $-0.022$ \\

    CN$_2$ & $= (0.22 \pm 0.06) \; \mathrm{log} \sigma + \gamma_{i}$ &
    $-0.390$ & 65 & 0.034 & $-0.376$ & 21 & 0.037 & $\;\;\:0.014$ &
    $0.009$ & $0.006$ & $-0.018$ \\

    log CaHK & $= (0.14 \pm 0.04) \; \mathrm{log} \sigma + \gamma_{i}$
    & $\;\;\:0.997$ & 65 & 0.048 & $\;\;\:1.061$ & 19 & 0.043 &
    $\;\;\:0.064$ & $0.012$ & $0.004$ & $-0.011$ \\

    CN3883 & $= (0.30 \pm 0.04) \; \mathrm{log} \sigma + \gamma_{i}$ &
    $-0.431$ & 65 & 0.051 & $-0.395$ & 15 & 0.075 & $\;\;\:0.036$ &
    $0.020$ & $0.008$ & $-0.026$ \\

    log G4300 & $= (0.14 \pm 0.08) \; \mathrm{log} \sigma +
    \gamma_{i}$ & $\;\;\:0.403$ & 65 & 0.051 & $\;\;\:0.437$ & 20 &
    0.058 & $\;\;\:0.034$ & $0.014$ & $0.004$ & $-0.024$ \\

  \enddata

  \tablenotetext{a}{Slope adopted from Kelson \etal (2001)}

  \tablecomments{(1) Scaling relation. (2) Zero point for the
  low-redshift sample. (3) Number of galaxies in the low-redshift
  sample. (4) rms in the $y$ direction of the low-redshift
  sample. (5,6,7) As 3,4,5 for the \rxj \ sample. (8) Zero point
  difference, $\gamma_{i \: \mathrm{\rxj}} - \gamma_{i \:
  \mathrm{low-}z}$. (9) The statistical uncertainties are calculated
  as
  \[
  \sigma_{\Delta \gamma \: i} =
  (\mathrm{rms}_{\mathrm{low-}z}^{2}/N_{\mathrm{low-}z} +
  \mathrm{rms}_{\mathrm{\rxj}}^{2}/N_{\mathrm{\rxj}})^{0.5}.
  \]
  (10) Systematic uncertainties on $\Delta \gamma \: i$, derived as
  0.026 times the coefficient of $\log \sigma$, based on the
  systematic uncertainties in $\log \sigma$ for the low-redshift
  sample. See J05 for more details. (11) Zero point difference as
  predicted by the relations in Table~\ref{tab:mod} and pure passive
  evolution.}

\end{deluxetable*}\nocite{kelson01,jorgensen05}

\begin{figure*}

  \begin{center}

    \epsscale{1.2}
    \plotone{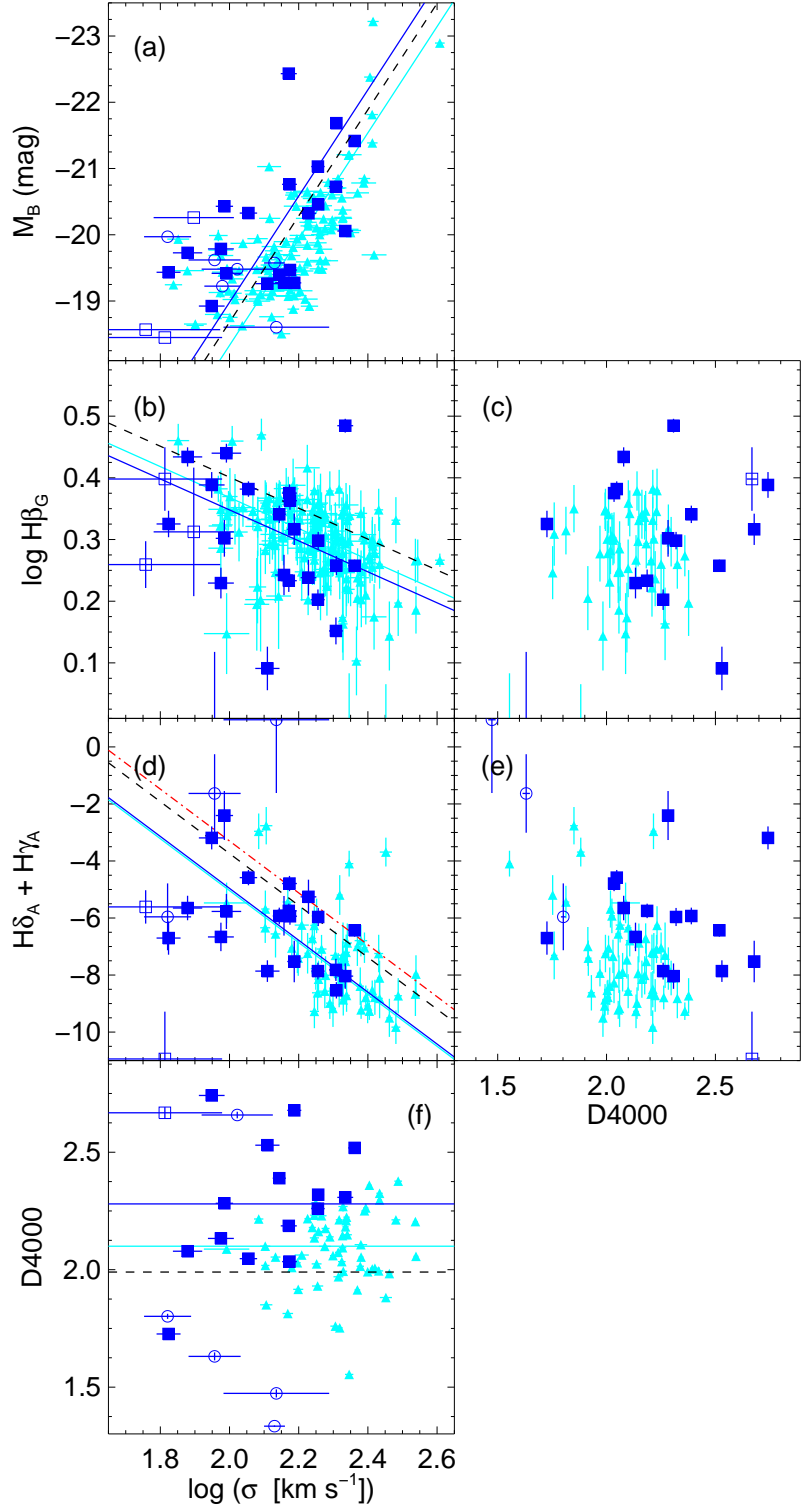}

    \figcaption{Scaling relations for age-dependant indices. (a) is the
    Faber-Jackson relation for \rxj. Triangles are galaxies from the
    low-redshift comparison sample. The light blue solid line is the
    linear fit to these points as described in the text. Boxes are
    galaxies in \rxj \ with no emission lines, circles are
    emission-line galaxies. The dark blue solid line is the fit to the
    non-emission-line galaxies in \rxj \ preserving the low-redshift
    slope (see text). Those points excluded from the fit are plotted
    as open boxes. The relationship expected for passively-evolving
    galaxies at $z = 0.28$, with a formation redshift of \zform $= 2$
    is indicated by the dashed line. The red dot-dashed line is the
    relationship found by Kelson \etal (2001) for a cluster of
    galaxies at $z=0.33$.\label{fig:avs}}

  \end{center}

\end{figure*}\nocite{kelson01}

\begin{figure*}

  \begin{center}

    \epsscale{0.85}
    \plottwo{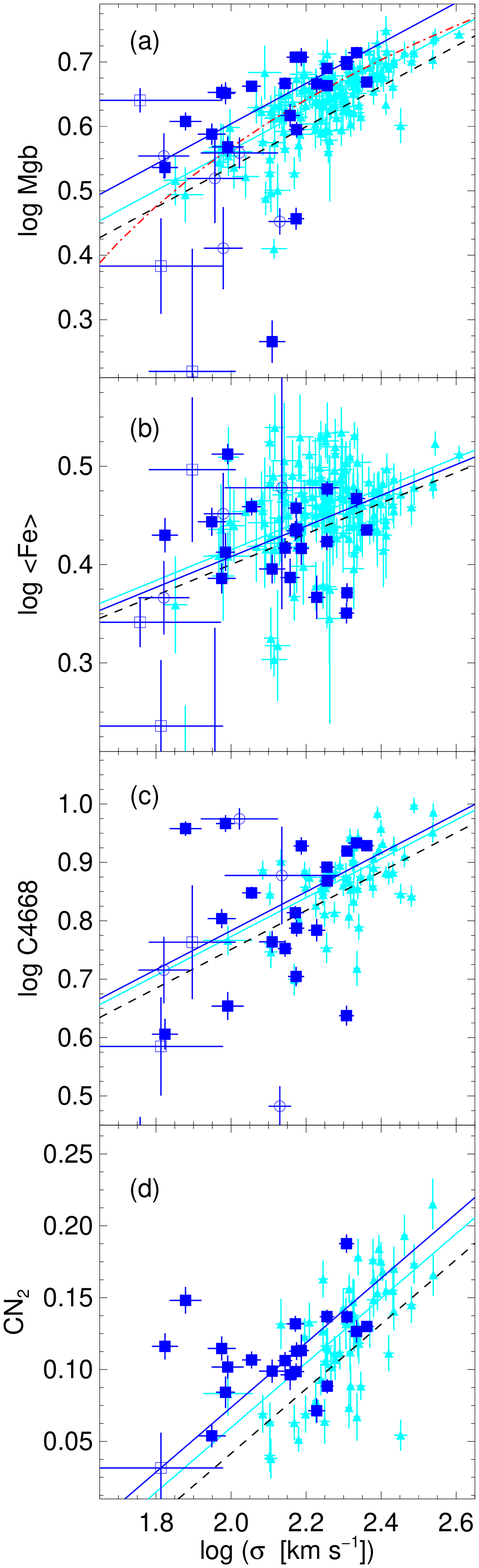}{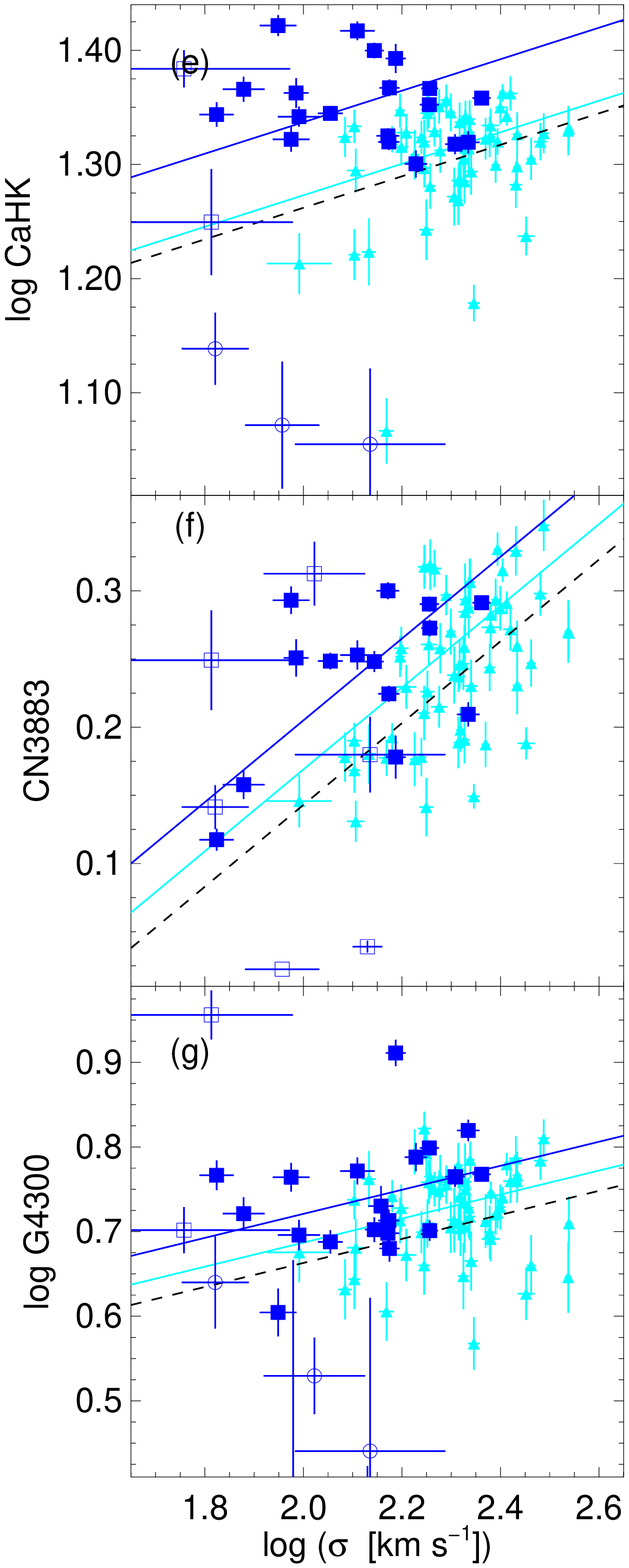}

    \figcaption{Scaling relations for the metal indices. The red
    dot-dashed line in (a) represents the relationship found by
    Colless \etal (1999) for cluster galaxies at $0.02 < z < 0.05$.
    All other plotting symbols and lines are as described in
    Figure~\ref{fig:avs}.\label{fig:1vs}}

  \end{center}

\end{figure*}\nocite{colless99}

We determine the scaling relations by fitting a linear relation to the
local galaxies. The corresponding relation for \rxj \ is
quantified as the median offset of the galaxies in \rxj, preserving
the slope. Linear fits to the low-redshift sample are made by
minimising the sums of the absolute residuals in the direction
perpendicular to the slope. Uncertainties in the slopes are calculated
using a bootstrap method. The zero points are derived by fitting to
the median of the measurements. This method has been shown to be very
robust to outliers. We quantify the random uncertainties in the zero
points ($\sigma_{\Delta\gamma \: i}$), and also the uncertainty due to
possible systematic effects in the determination of the velocity
dispersion ($\sigma_{\mathrm{sys \: i}}$). The latter quantity is
based on the systematic uncertainty of 0.026 in $\log \sigma$, as
derived in J05. The total uncertainty in the relation is equal to
$\sigma_{\Delta \gamma \: i} + \sigma_{\mathrm{sys \: i}}$. For \HdgA
\ vs. log $\sigma$ we adopt the slope, determined by Kelson \etal
(2001)\nocite{kelson01}, of CL1358+62 at $z=0.33$. The zero point in
this case is the median value of \HdgA \ of the low-redshift galaxies.

When determining the relationships for \rxj, we exclude those cluster
members with obvious emission-lines, indicated in
Table~\ref{tab:lin}. We also exclude the galaxies with $\sigma_{\log
\sigma} > 0.1$. The comparison in this section is therefore between E
and S0 galaxies in the local sample and galaxies within 0.2 mag of the
red sequence in \rxj.

Figure~\ref{fig:avs} plots age-dependant observables against velocity
dispersion. Figure~\ref{fig:avs}a is the FJ relation between
rest-frame $B$-band magnitude and velocity dispersion. A cursory
examination of the distribution of velocity dispersions in each sample
indicates that \rxj \ lacks high-$\sigma$ galaxies. This can be seen
with most emphasis in Figure~\ref{fig:avs}b and Figures~\ref{fig:1vs}a
and \ref{fig:1vs}b. Furthermore, there are only 2 red-sequence
galaxies within 1.5 mag of the BCG magnitude not selected for
spectroscopy (see Figure~\ref{fig:cmd}). This suggests that the
paucity of high-$\sigma$ galaxies is real, not a function of our
spectroscopic selection. The FJ relation shows that the BCG in \rxj \
has conspicuously low velocity dispersion for its luminosity. The
scatter in the \rxj \ points is larger than that of the control sample
and the offset is consistent with that expected from passive evolution
of the galaxy population within the errors. A more accurate measure of
the luminosity evolution can be made when the parameters are better
constrained by the FP, and hence will have to wait until \hst \
observations of the surface brightnesses and effective radii are
analysed (Barr et al., in preparation).

Figure~\ref{fig:avs}b-f shows scaling relations for \Hbetaem \ and
\HdgA \ and D4000.  The Balmer line indices show marginal
inconsistency with passive evolution, ($1.4\sigma$ and $1.1\sigma$
respectively).  The strength of the 4000\AA \ break is uncorrelated
with velocity dispersion (a Kendall's $\tau$ correlation test
indicates that there is no significant correlation in either data
set). The values for \rxj \ are also significantly offset from those
in the comparison sample; a two-sided Kolmogorov-Smirnoff test
indicates that there is a $<1\%$ chance that the two sets of D4000
measurements are drawn from the same parent distribution. Furthermore,
passive evolution of galaxies in a cluster would predict that D4000 is
stronger at lower redshift, in contrast to
Figure~\ref{fig:avs}d. Using the relationship in Table~\ref{tab:sca},
we estimate that the median value of D4000 for non-emission-line
galaxies is stronger than expected from the passive-evolution model at
greater than 4 times the uncertainty. Because the 4000\AA \ break is
predicted to be stronger for older stellar populations, there are no
models which allow for this offset to occur through passive evolution.

Figure~\ref{fig:1vs} shows the scaling relations for metal indices. In
the cases of \fe \ and C4668 the difference in the populations can be
explained by age differences alone. However, in all other cases the
relation is offset significantly in the opposite direction to that
expected from passive evolution. We can quantify the average,
normalised offset for $N$ indices from passive evolution as,

\[ 
\Delta \gamma _{n} = \frac{1}{N} \sum_{i}^{N}
\frac{\mathrm{PE}_{i} - \Delta \gamma_{i}}{\mathrm{PE}_{i}}
\]

\noindent
where $\Delta \gamma_{i}$ is the offset of each index from the
comparison sample, and $\mathrm{PE}_{i}$ is the passive evolution
offset (see Table~\ref{tab:sca}). The uncertainties are then,

\[ 
\sigma_{\Delta \gamma \: n} = \frac{1}{N}
\sqrt{\sum_{i}^{N} \left ( \frac{\sigma_{\Delta \gamma \:
i}}{\mathrm{PE}_{i}} \right )^{2} }
\]

\noindent
and,

\[ 
\sigma_{\mathrm{sys} \: n} = \frac{1}{N} \sum_{i}^{N} \frac{
\sigma_{\mathrm{sys} \: i}}{|\mathrm{PE}_{i}|}
\]

\noindent
with the total uncertainty equal to $\sigma_{\Delta \gamma \: n} +
\sigma_{\mathrm{sys} \: n}$ as before. The metal indices combined
(excluding \fe \ and C4668) are stronger than predicted by the passive
evolution model, with 5 times the uncertainty. This result is not
driven by a single measurement as it still holds (at 3.7 times the
uncertainty) if we exclude CaHK, the most discrepant index. If \fe \
and C4668 are included, the enhancement remains at 3.9 times the
uncertainty.

With the exception of \fe \ and CaHK, the rms scatter in the relations
is higher for \rxj \ than the low-redshift comparison sample. This is
not due to higher measurement errors, as the median uncertainty for
the line indices in the \rxj \ sample is generally lower than that of
the comparison sample.

\subsection{Age-metallicity-\afe \ indicators}

Figure~\ref{fig:xvm} shows the visible line indices and D4000 against
one another. Model grids from Thomas \etal (2003a,
2004)\nocite{thomas03,thomas04} are overlaid.

Figure~\ref{fig:xvm}b plots \Hbetaem, an age indicator, against the
quantity $[\mathrm{MgFe}]^{\prime}$, constructed by Thomas \etal
(2003a)\nocite{thomas03}\footnote{$[\mathrm{MgFe}]^{\prime} = \sqrt{
\mathrm{Mg}b \cdot ( 0.72 \: \mathrm{Fe5270} + 0.28 \: \mathrm{Fe5335}
) }$ } to be an index independent of \afe. We can infer from this plot
that the age and metallicity distributions of the low-redshift
comparison sample and \rxj \ are similar. Figure~\ref{fig:xvm}g
complements this analysis in that it probes $[\alpha/\mathrm{Fe}]$
space where age and metallicity are degenerate. In this plane it can
be seen that \afe \ is higher in \rxj \ than in the local galaxies.

\begin{figure*}

  \begin{center}

    \epsscale{1.1}
    \plotone{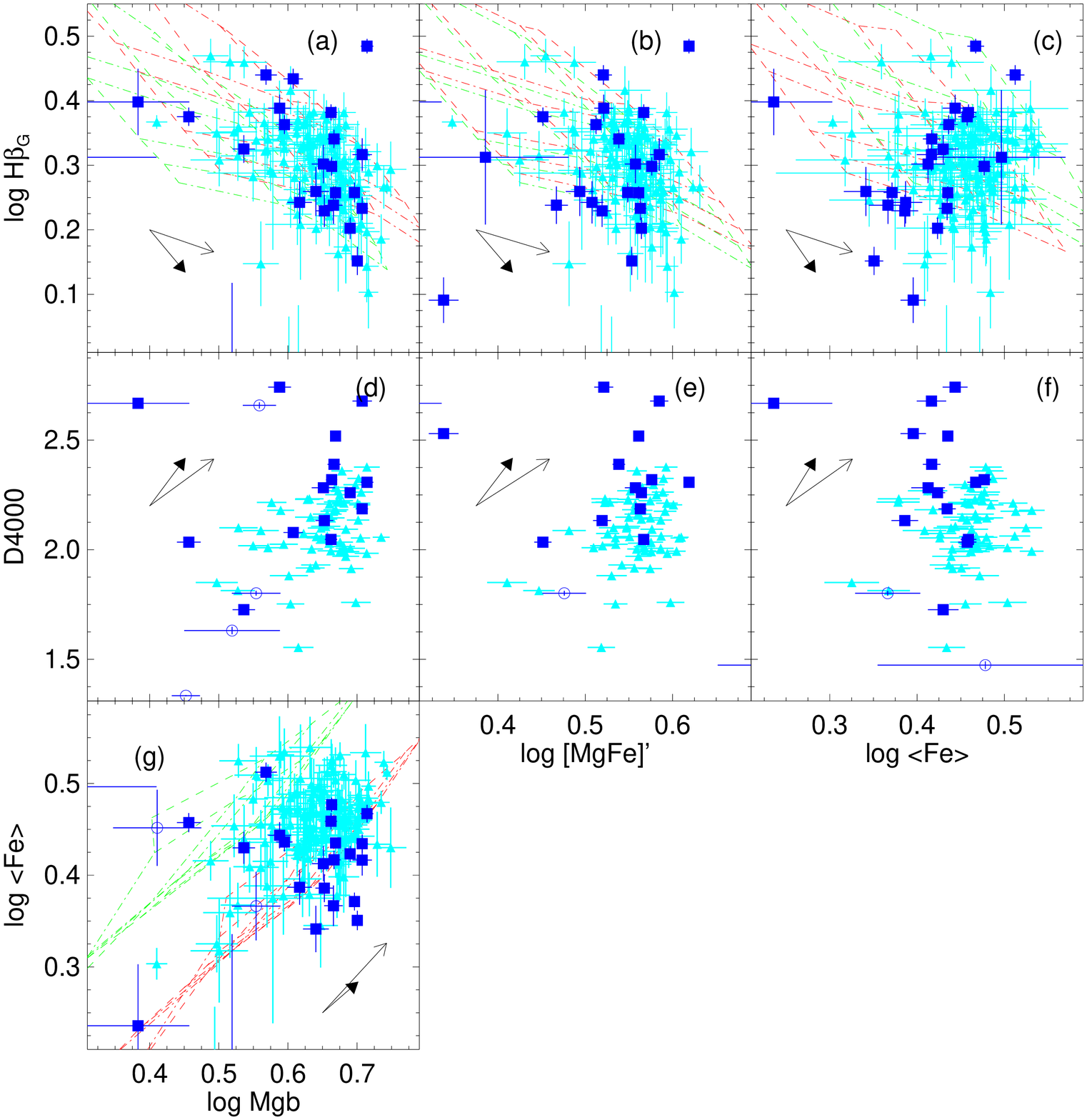}

    \figcaption{Visible line indices and D4000 versus each other. The
    plotting symbols are the same as those in
    Figure~\ref{fig:avs}. Where available, we also overlay the grids
    of Thomas \etal (2003a, 2004), color-coded by \afe \ ratio. Green
    represents \afe \ = 0.0 and red \afe \ = 0.3. Dashed lines are
    lines of constant metallicity with values of [M/H] = -0.30, 0.00,
    0.35 0.67. The dot-dashed lines indicate metallicity changes at
    constant age with 1, 2, 4, 8, 11 and 15 Gyr. The arrows show the
    approximate changes in the indices for a change of $\Delta \log
    \mathrm{age} = 0.3$ (solid arrow) and $\Delta \mathrm{[M/H]} =
    0.3$ (open arrow). The ordinate axis in panels (b) and (e) is
    designed such that quantities are independent of
    \afe.\label{fig:xvm}}

  \end{center}

\end{figure*}\nocite{thomas03}

We now derive values for luminosity-weighted mean age, metallicity and
\afe \ from \Hbetaem, \mgb \ and \fe \ using the models of Thomas
\etal (2003a)\nocite{thomas03}. We use \Hbetaem, \mgb \
and \fe \ because these are generally the best studied indices (\eg
J\o rgensen 1999; Trager \etal 2000)\nocite{jorgensen99,trager00b},
and are consistent with observations of globular clusters, \ie systems
that can truly be thought of as SSPs. However, caution must be
exercised, and results cannot be interpreted as absolute values of the
physical parameters. This can be seen from Figure~\ref{fig:xvm}, and
in particular the grids which involve \Hbetaem \ where the Thomas
models appear to overpredict the value of the index. In many cases,
points from both the high- and low-redshift sample lie off the model
grids. Because of this, the absolute luminosity-weighted mean age of a
particular galaxy returned may be improbably old. We therefore
concentrate on the median relative difference in luminosity-weighted
mean age, metallicity and \afe \ between the samples and use these,
rather than absolute values of the physical properties, to interpret
our results.

We fit the three indices simultaneously, and linearly interpolate
between points on the model grid. We also extrapolate linearly beyond
the edge of the grid. The values of age, [M/H] and \afe \ which
minimise $\chi^2$ in \Hbetaem, \mgb \ and \fe \ are adopted. Errors
are estimated by doing the same at the 6 extreme points of the error
ellipsoid defined by the uncertainties on the indices. Half the
maximum of the differences at these 6 points is used at the
uncertainty estimate. This representation of the uncertainty is
consistent with the results of Monte-Carlo simulations of the model
inversion. In these simulations, we use a subset of galaxies in the
low-redshift sample, and vary the errors in \Hbetaem, \mgb \ and \fe \
with Poissonian probability. The standard deviation of values in
luminosity-weighted mean age, [M/H], and \afe \ more closely match the
errors derived using the maximum differences than the mean or median
differences of the points in the error ellipsoid.

\begin{figure*}

  \begin{center}

    \epsscale{1.1}
    \plotone{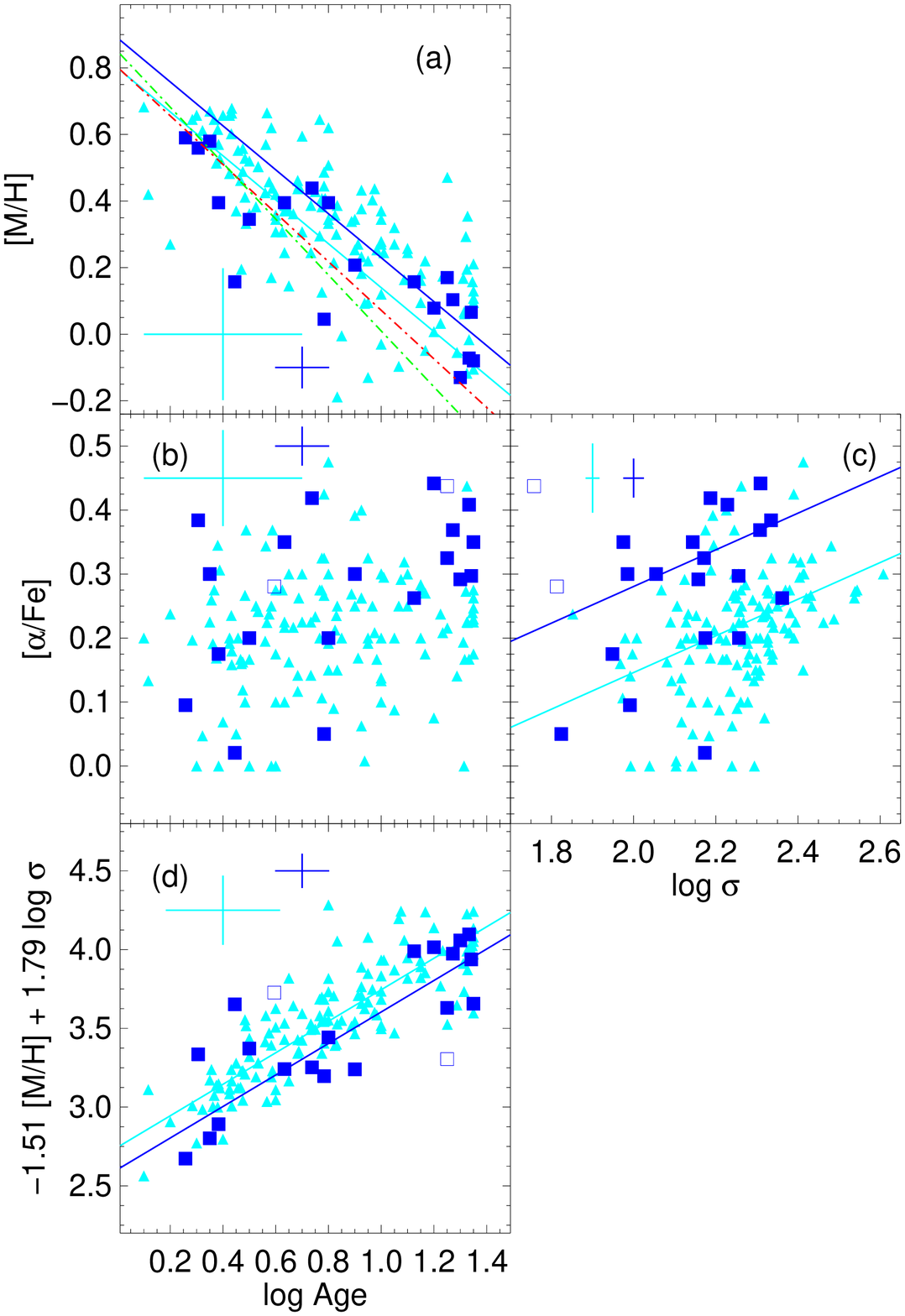}

    \figcaption{Derived relative physical attributes of cluster
    galaxies plotted against each other. Triangles are galaxies in the
    local sample, while squares are non-emission line galaxies in
    \rxj, open symbols denoting galaxies excluded from the fits
    described in the text. The error bars show the median error in
    each sample. The fits to the points are shown as solid lines. In
    (a) the lines represent the age-metallicity relationship for
    galaxies with $\log \sigma = 2.2$ in each sample; relations found
    by J{\o}rgensen (1999) and Trager \etal (2000) are denoted by the
    dot-dashed lines, green and red respectively. Panel (d) shows
    edge-on view of the log age -- [M/H] -- $\log \sigma$
    plane.\label{fig:sva}}

  \end{center}

\end{figure*}\nocite{trager00b,jorgensen99}

The results are shown in Figure~\ref{fig:sva}. Correlations between
age, [M/H] and velocity dispersion, and \afe \ and velocity dispersion
\cite{jorgensen99,trager00b} are documented for galaxies at low
redshift. We quantify the \{log age, [M/H], log $\sigma$\} plane by
extending the method used to fit our scaling relations to three
dimensions. An equation of the form,
$$\log \mathrm{age} = (-1.51 \pm 0.01) [\mathrm{M}/\mathrm{H}] + (1.79
\pm 0.23) \log \sigma$$
\noindent
$$- (2.75 \pm 0.02)$$
\noindent
is fit to the low-redshift data by minimising absolute residuals
perpendicular to the plane. This plane is broadly consistent with that
of J{\o}rgensen (1999)\nocite{jorgensen99} and Trager \etal
(2000)\nocite{trager00b} (see Figure~\ref{fig:sva}). As with our
previous fits for \rxj \ data points, we preserve the slope of this
relation and find a median offset of $0.14 \pm 0.07$ in log age. This
means that at a given velocity dispersion and metallicity, the
luminosity-weighted mean ages of galaxies in \rxj \ are $0.14 \pm
0.07$ dex {\em older} than they are in the low-redshift sample. We
address the somewhat counter-intuitive notion that galaxies at higher
redshift can have older stellar populations than those in the local
universe in \S\ref{sec:dis}. We also fit the scaling relation in \afe
\ vs. $\log \sigma$, using the method outlined in \S\ref{sec:vds}. The
galaxies in \rxj \ have a median value of \afe \ which is $0.14 \pm
0.03$ higher than the comparison sample.

Figure~\ref{fig:sva} shows these relationships, using a representative
value of $\log \sigma = 2.2$ in the age--metallicity diagram. It also
shows a side-on view of the \{log age, [M/H], log $\sigma$\} plane in
Figure~\ref{fig:sva}d. In this view it appears evident that there is a
bimodality in the luminosity-weighted mean ages of galaxies in
\rxj. There is no spatial clustering or coincidence of redshifts for
galaxies with log age $> 1.0$, although they do include 5 of the 6
brightest galaxies in the spectroscopic sample. The six galaxies in
\rxj \ which appear to lie on the $\log \sigma -$\afe \ relation found
at low redshift (those with $0.05 <$ \afe \ $<0.28$) display no
obvious spatial clustering.

\section{Discussion}
\label{sec:dis}

In this section we first assess the derived offsets in our scaling
relations and their consequences for the passive evolution model. We
then discuss the evidence for stellar populations with older
luminosity-weighted mean ages and larger \afe \ ratios in
\rxj. Finally we speculate on scenarios which can give rise to the
inferred properties of galaxies in \rxj, and the evolution of such a
cluster to $z \sim 0$.

Of the scaling relations involving age indicators, the Faber-Jackson
relation is consistent with passive evolution while the Balmer line
indices are inconsistent only at the $1\sigma$ level. D4000 gives a
much stronger result: its values are inconsistent with passive
evolution at $>4$ times the uncertainty. D4000 has been invoked as an
age indicator (\eg Tremonti \etal 2001; Kauffmann \etal
2003\nocite{tremonti01,kauffmann03}), though metallicity is at least
as significant according to the relation given in Table~\ref{tab:mod}.

The scaling relations for \mgb, CN$_2$, CaHK, CN3883, G4300 in \rxj \
are offset in the direction opposite to that predicted by passive
evolution. When combined, this offset is 5 times the uncertainty. \fe
\ and C4668 are the only metal indices that are consistent with the
passive-evolution model.

For the pure passive evolution scenario to be correct, either our data
or the SSP models have to be wrong. Systematic errors in the velocity
dispersion have already been taken into account when arriving at the
above results. We consider systematics affecting the line indices
unlikely: the only scaling relations consistent with passive evolution
belong to those indices whose dependence on age is small relative to
that on metallicity or \afe. Furthermore, the scaling relations with
significant offset all indicate older ages. It is difficult to imagine
a systematic error that could produce such a consistent offset, as
some indices would have to get stronger and some weaker.

SSPs with non-solar \afe \ are yet to be widely tested or
used. Nevertheless their age dependencies are generally consistent
with those that employ solar \afe \ (\eg Vazdekis \etal
1997\nocite{vazdekis97}). We note that none of the results presented
in \S\ref{sec:vds} are changed if we use the model line indices in
Vazdekis \etal (1997)\nocite{vazdekis97} as our reference, as opposed
to those of Thomas \etal The inconsistency of the results with the
passive evolution model is thus not a consequence of our comparison
with the Thomas models.

The assertion that the \afe \ ratio in members of \rxj \ is enhanced
over the control sample makes use of the indices' dependence on \afe \
specific to the Thomas models. Even so, it is generally accepted that
\mgb \ is more sensitive to Mg than to Fe, and \fe \ is more sensitive
to iron than to $\alpha$ elements. The relative strength of \mgb \ in
Figure~\ref{fig:1vs}a and weakness of \fe \ in Figure~\ref{fig:1vs}b,
combined with the high ratio of the the \mgb \ index to \fe \ in
Figure~\ref{fig:xvm}g, indicates directly that \afe \ is enhanced in
\rxj. Even if \afe \ is not yet accurately modeled, we are forced to
conclude, albeit in a qualitative sense, that it is significantly
enhanced compared with the low-redshift sample.

By inverting the models for \Hbetaem, \mgb, and \fe, we calculate
that, at a given velocity dispersion and metallicity, galaxies in \rxj
\ have luminosity-weighted mean ages $0.14 \pm 0.07$ dex older than
the comparison sample. A similar analysis indicates that \afe \ in
stellar populations in \rxj \ is $0.14 \pm 0.03$ greater than at low
redshift.

Three indices are used to obtain these offsets. In order to give
preference to this interpretation over the pure passive evolution
model, we must test these values against all scaling relations, and in
particular those which show consistency with passive evolution. We
find that the offsets for all indices in Table~\ref{tab:mod} modeled
by Thomas are consistent with $\Delta \log \mathrm{age} = 0.14$ and
$\Delta$\afe \ $ = 0.14$. This check is more difficult to make for the
CMD, the FJ relation, D4000, CaHK, and CN3883 for which the models
assume solar abundance ratios. In these instances we gauge whether the
\afe \ dependence required, given the difference in log age and the
equations in Table~\ref{tab:mod}, is feasible. CN3883 requires only a
weak correlation with \afe, similar to the related index CN$_2$. If we
assume that the dependence is the same, then CN3883 is consistent with
our derived offsets.

Only about half of the average difference in the 4000\AA \ break can
be explained by $\Delta \log \mathrm{age} = 0.14$. In order to be
consistent with our \afe \ offset, D4000 would have to become stronger
with increasing $\alpha$-element abundance ratio. For our low-redshift
sample, we find that D4000 is weakly positively correlated with
\afe. J05 also found evidence for such a correlation. The greater
average strength of the 4000\AA \ break therefore appears
qualitatively consistent with our estimated offsets in age and
abundance ratio.

A greater problem is presented by the FJ relation and the CMD. Stellar
populations with $\Delta \log \mathrm{age} = 0.14$ are expected to be
$\sim 0.3$ mag fainter than the comparison sample. For our FJ relation to
be consistent with our derived offsets, the $B$-band mass-to-light
ratio of a galaxy would have to decrease with increasing
$\alpha$-element abundance ratio. Furthermore, for the CMD to remain
consistent, either the observed \rfil-band magnitude would have to
brighten, or stellar populations would have to become bluer, with
increasing \afe.  There are no models which predict the behavior of
magnitude and color vs. \afe. However, Thomas \& Maraston
(2003)\nocite{thomas03b} indicate that blue luminosity increases with
increasing \afe, which would work to push older stellar populations in
the CMD and FJ relations back toward consistency with the passive
evolution predictions. As far as the CaHK index is concerned, we
require that it becomes stronger with increasing \afe. This seems
sensible, given that Ca is itself an $\alpha$-element. However, Ca has
been found to be underabundant compared with other $\alpha$-elements
in local early-type galaxies \cite{thomas03c}. There are also no SSP
models which estimate its dependence on \afe. Because of this rather
complex relationship, we note that the CaHK index is incompatible with
the passive evolution model but can provide no evidence to support our
alternative model.

The scaling relations for indices modeled with \afe \ are consistent
with $\Delta \log \mathrm{age} = 0.14$ and $\Delta$\afe \ $ =
0.14$. Furthermore, those scaling relations (including the CMD and the
FJ relation) involving observables whose dependence on \afe \ is not
known display, qualitatively, the behavior expected of older stellar
populations with enhanced abundance ratios. We conclude that the
combined scaling relations show that galaxies in \rxj \ have greater
\afe \ than those at $z \sim 0$ and cannot evolve into them by passive
evolution.

Two questions immediately arise from stellar populations in \rxj \
with higher \afe \ than those in low-redshift clusters. Firstly, how
do cluster galaxies come to have high \afe \ ratios? Secondly, is
there a mechanism by which stellar populations in \rxj \ can reduce
their \afe \ and become like those at $z \sim 0$?

Enhanced \afe \ ratios can be produced by a short episode of star
formation which is curtailed before the lower-mass stars can produce
significant Fe-peak elements through Type I supernovae. In a scenario
such as this the luminosity-weighted mean ages of galaxies would be
higher than expected because ongoing star-formation is absent. Such an
episode would have to be quenched within $\sim 1$ Gyr to prevent the
products of Type Ia SNe contaminating a subsequent generation of
stars. This can be mitigated to some extent if the star formation
episode is violent enough to produce a top-heavy IMF which will give a
higher yield of $\alpha$ elements per unit luminosity (\eg see Worthey
\etal 1992)\nocite{worthey92}.

If \rxj \ formed rapidly in a cluster merger, the stripping of gas
from galaxies in such an event could be used to explain the rapid
quenching of star-formation. The large velocity dispersion of \rxj \
might thereafter inhibit the merging of cluster members and the
associated decrease of \afe \ and luminosity-weighted mean age. This
could also be used to explain the lack of a dominant galaxy. It does
not, however, explain the rather ordinary fraction of blue galaxies,
which we might expect to be higher than normal because of the absence
of merging. Values of \afe \ for stellar populations have yet to be
estimated in a large number of distant clusters, so it is unclear
whether \rxj \ is unusual in this respect. However, the difference in
the implied luminosity-weighted mean ages does mean that any
evolution from the stellar populations in \rxj \ to those in
our comparison sample cannot be by pure passive evolution.

No substructure is detectable in \rxj \ based on our data, and there
is no evidence for an excess of star-forming, or post-starburst
galaxies.  However, its position on the $B_{gc} - \log \sigma$ and
$L_{X} - \log \sigma$ suggests a large cluster velocity dispersion for
its luminous mass. In terms of its position on the $B_{gc}$ vs. $\log
\sigma$ plot, \rxj \ most resembles MS~1455+22 at $z=0.26$, (see Yee
\& Ellingson 2003)\nocite{yee03}. However, MS~1455+22 has a very large
cD galaxy, strong cooling flow, and a very low fraction of blue
galaxies. In local terms, the cluster closest to \rxj \ in $B_{gc} -
\log \sigma$ is Abell 85 at $z=0.055$ \cite{yee99}. Abell 85 is in the
early stages of a merger and has a moderately strong cooling flow
associated with a cD galaxy \cite{kempner02,durret03}. It seems
unlikely, therefore, that there is much correspondence between \rxj \
and either of these clusters. It is possible that \rxj \ has no
analogy in the low-redshift universe, or that local surveys cannot
discover massive, underluminous clusters. This seems highly improbable
as such clusters should still be detectable to X-ray
observatories. \rxj \ could have a significantly lower X-ray
luminosity than inferred from the \rosat \ observation if there is a
contribution to the X-ray flux from obscured AGNs which \rosat \
cannot resolve spatially. \chandra \ or \xmm \ observations would be
invaluable in resolving this question. However, the problem of the
prodigious cluster velocity dispersion remains whether or not there is
significant X-ray flux in AGNs. There remain hints that the centre of
\rxj \ is not yet fully virialized; for example, the lack of a
dominant galaxy, the velocity offset of the brightest member from the
cluster centre and the distance between the brightest and second
brightest members. It is clear that \rxj \ is an unusual cluster of
galaxies. Further study, and in particular high-resolution X-ray
observations, will be invaluable in discerning its nature.

Whatever its history, it is difficult, if not impossible, to imagine a
scenario by which the galaxies in \rxj \ can evolve, in the $\sim 3$
Gyr available, from their state at $z = 0.28$ to that of our
comparison sample. Star formation (which is not present at $z = 0.28$)
would have to occur within the early-type population to decrease their
luminosity-weighted mean ages. This would have to occur gradually in
order to avoid enhancing the \afe \ even further, which presumably
rules out a cluster merger event. Luminosity-weighted mean ages and
abundance ratios could be reduced if there is a great deal of merging
between elliptical and disky galaxies. However, the lack of an excess
of blue galaxies and a high cluster velocity dispersion argues against
an unusual amount of merging being able to take place. Furthermore, it
has been argued that the blue fraction is reduced mainly by the ageing
of stellar populations \cite{balogh99}. It remains difficult to devise
a mechanism in clusters by which a lot of merging, but little star
formation, takes place.

\section{Conclusions}
\label{sec:con}

We present GMOS-N broad-band and MOS observations of \rxj, an X-ray
luminous cluster of galaxies at $z=0.28$. Redshifts, central velocity
dispersions, and absorption-line indices for 43 objects are
derived. Of the spectroscopic sample, 30 galaxies are members of
\rxj. We calculate the cluster velocity dispersion as $1278 \pm 134$
km s$^{-1}$ and find no sign of substructure. Our broad-band
observations show \rxj \ to be a poor cluster with a blue fraction
consistent with other clusters at $z \sim 0.3$. It appears to have a
very large velocity dispersion for both its richness and X-ray
luminosity.
 
The $($\gfil$-$\rfil$)$ vs. \rfil \ red sequence for \rxj \ is
consistent with a passively evolving population of early-type galaxies
formed at \zform $\simeq 2$.

We have established scaling relations between the absolute $B$-band
magnitude and central velocity dispersion (the Faber-Jackson
relation), as well as relations between absorption line indices and
central velocity dispersion for galaxies in \rxj. These are compared
with the low-redshift sample and predictions from single stellar population
models. The Faber-Jackson relation is in agreement with a pure passive
evolution model and the Balmer indices (\Hbetaem, \HdgA) are in
marginal disagreement. The scaling relations for \mgb, CN$_2$, CaHK,
CN3883, and G4300 indices, when taken together, are offset from the
pure passive evolution model with five times the uncertainty.

We use the models of Thomas \etal and the measured indices \{\Hbetaem,
\mgb, \fe\} to compute the relative differences in luminosity-weighted
mean age, metallicity and $\alpha$-element abundance ratio between
stellar populations in \rxj \ and the low-redshift comparison
sample. We find that for a particular metallicity and velocity
dispersion, \rxj \ has stellar populations with luminosity-weighted
$\log \mathrm{age} = 0.14 \pm 0.07$ older than our comparison
sample. We also show that galaxies in \rxj \ have \afe \ $0.14 \pm
0.03$ greater than the local cluster galaxies. All scaling relations
are consistent with these values, and so we consider this scenario
more likely than one in which stellar populations evolve passively. We
speculate that these enhancements were caused by rapid bursts of star
formation that were subsequently curtailed.
  
\rxj \ appears to be without a counterpart in the local
universe. Current models of cluster evolution are unable to provide a
path by which \rxj \ can evolve into our low-redshift comparison
sample. We believe that the morphology of the cluster gas in \rxj \
will provide clues as to its formation history. X-ray observations
with high spatial resolution will be required for this analysis.

\acknowledgments

The GMOS instrument, commissioning and System Verification teams are
thanked for their effort in making GMOS an efficient Gemini facility
instrument. The observations used in this paper were obtained as part
of the System Verification observations, program ID
GN-2001B-SV-51. RLD is grateful for the award of a PPARC Senior
Fellowship (grant number PPA/Y/S/1999/00854). Based on observations
obtained at the Gemini Observatory, which is operated by the
Association of Universities for Research in Astronomy, Inc., under a
cooperative agreement with the NSF on behalf of the Gemini
partnership: the National Science Foundation (United States), the
Particle Physics and Astronomy Research Council (United Kingdom), the
National Research Council (Canada), CONICYT (Chile), the Australian
Research Council (Australia), CNPq (Brazil) and CONICET (Argentina).


\appendix

\section{GMOS-N data reduction}
\label{sec:dre}

\subsection{Imaging data}

Broad-band images in \gfil, \rfil \ and \ifil \ are bias-subtracted
and then flatfielded using twilight skyflats. Fringing is significant
(at about 1\% of the sky background) in the \ifil \ band, while
scattered light needs to be removed from all frames. Scattered light
and fringe frames are created by combining images in each filter with
the brightest objects masked out. These are fitted with a smooth
surface before being subtracted from each image. In the case of the
\ifil-band, where fringing is significant, the fringe frame is
median-smoothed with a \arcsd{1}{8}$\times$\arcsd{1}{8} box before
subtraction. Images are then registered and combined rejecting cosmic
rays and bad pixels.

\subsection{Photometric calibration}
\label{sec:crf}

Flux calibrations are accomplished using observations of standard
stars on UT 2001 October 21 under photometric conditions. Science
exposures from other nights were normalised to the observations taken
on UT 2001 October 21. The photometry is corrected for the effect of
galactic extinction using the prescription of Cardelli \etal
(1989)\nocite{cardelli89} based on the measurement of $A_{B}
= 0.292$ toward \rxj \ of Schlegel \etal
(1998)\nocite{schlegel98}. Values of galactic extinction in each
filter are $A_{g^{\prime}} = 0.243$, $A_{r^{\prime}} = 0.189$ and
$A_{i^{\prime}} = 0.150$.

Standard magnitudes are derived using the relation,

\[
m_{\mathrm{std}} = m_{\mathrm{zero}} + \Delta m_{\mathrm{zero}} - 2.5
\log (N/t) - k(airmass - 1)
\]

\noindent
where $airmass$ is the mean airmass of the observation and $k$ is the
mean atmospheric extinction at Mauna Kea. Magnitude zero points and
the dependence of $\Delta m_{\mathrm{zero}}$ on color are given in
Table~\ref{tab:col}.

\begin{deluxetable}{cccccc}

  \tablewidth{0pt} 
  \tablecolumns{6}
  \tablecaption{Photometric zero points and color terms for
  photometric calibration\label{tab:col}}

  \tablehead{\colhead{filter} & \colhead{$m_{\mathrm{zero}}$} &
  \multicolumn{4}{c}{$\Delta m_{\mathrm{zero}}$} \\ \colhead{} &
  \colhead{} & \colhead{rms} & \colhead{Color term fit} &
  \colhead{rms (fit)} & \colhead{Color interval} \\ \colhead{} &
  \colhead{(1)} & \colhead{(2)} & \colhead{(3)} & \colhead{(4)} &
  \colhead{(5)} }

  \startdata

    \gfil & $27.88 \pm 0.01$ & $0.044$\tablenotemark{a} & $(0.066 \pm
    0.002) ($\gfil$ - $\rfil$ ) - (0.037 \pm 0.002)$ & $0.034$ &
    $-0.55 \leq ($\gfil$ - $\rfil$ ) \leq 2.05$ \\

    \rfil & $28.15 \pm 0.01$ & $0.045$\tablenotemark{b} & $(0.042 \pm
    0.004) ($\rfil$ - $\ifil$ ) - (0.011 \pm 0.002)$ & $0.043$ &
    $-0.35 \leq ($\rfil$ - $\ifil$ ) \leq 2.20$ \\

    \ifil & $27.86 \pm 0.01$ & $0.054$\tablenotemark{b} & $(0.063 \pm
    0.005) ($\rfil$ - $\ifil$ ) - (0.013 \pm 0.002)$ & $0.050$ &
    $-0.35 \leq ($\rfil$ - $\ifil$ ) \leq 2.20$ \\
  
  \enddata

  \tablecomments{(1) Photometric zero point, (2) rms of $\Delta m$,
  equivalent to the expected uncertainty on the standard calibration
  if the color terms are ignored, (3) linear fits to the color
  terms, (4) rms of the linear fits, (5) color interval within which
  the linear fit applies.}

  \tablenotetext{a}{$-1.10 \leq ($\gfil $-$ \ifil$) \leq 3.05$}
  \tablenotetext{b}{$-0.7 \leq ($\rfil $-$ \ifil$) \leq 2.5$}

\end{deluxetable}
  
Our low-redshift comparison sample of galaxies is calibrated in
rest-frame $B$. The exact transformation from \rfil\ifil \ photometry
to this system depends on the redshift. Full details of the
methodology of rest-frame calibrations for clusters in the \pname \
will be given in a future paper outlining the GMOS-N photometry for
the project. The transformation from \rfil\ifil \ to
$B_{\mathrm{rest}}$ at the redshift of \rxj \ ($z=0.280$) is,

\[ 
B_{\mathrm{rest}} = i^{\prime} + 0.4753 + 1.6421\,( r^{\prime} -
i^{\prime} ) - 0.0253\,( r^{\prime} - i^{\prime} )^2
\]

\noindent 
and in all cases, the absolute $B$-band magnitude is derived as

\[
M_{B} = B_{\mathrm{rest}} - DM(z) + 2.5 \log ( 1 + z )
\]

\noindent
where the distance modulus ($DM(z)$) for the cluster redshift in our
adopted cosmology, is 40.78.

\subsection{Spectroscopic data}
\label{sec:sdr}

Table~\ref{tab:pho} gives the positions, magnitudes and colors of
galaxies in the spectroscopic sample. 

\begin{deluxetable}{rcccccccc}

  \tablewidth{0pt} 
  \tablecolumns{8}
  \tablecaption{Photometric properties of the spectroscopic
  sample\label{tab:pho}}

  \tablehead{\colhead{ID} & \colhead{RA (J2000)} & \colhead{Dec
  (J2000)} & \colhead{$g'$} & \colhead{$r'$} & \colhead{$i'$} &
  \colhead{($g'-r'$)} & \colhead{($g'-i'$)} & \colhead{B/D}}

  \startdata

         1 & 1:42:09.11 & 21:33:23.8 &  19.20 &  18.02 &  17.50 &    1.57 &    2.17 &        B   \\
    22 & 1:42:08.68 & 21:33:22.6 &  20.03 &  18.97 &  18.47 &    1.62 &    2.22 &        B   \\
    44 & 1:42:10.37 & 21:33:32.2 &  22.62 &  21.36 &  20.87 &    1.32 &    1.85 &  \nodata   \\
    88 & 1:42:09.21 & 21:33:14.0 &  18.48 &  19.59 &  19.40 &    1.14 &    1.57 &        D   \\
   116 & 1:41:54.76 & 21:33:08.5 &  21.00 &  20.19 &  19.85 &    0.92 &    1.35 &        D   \\
   128 & 1:42:07.37 & 21:33:02.1 &  20.79 &  19.38 &  18.88 &    1.50 &    2.08 &        B   \\
   173 & 1:41:56.30 & 21:32:54.9 &  21.08 &  20.20 &  19.80 &    1.02 &    1.50 &  \nodata   \\
   205 & 1:42:04.40 & 21:32:39.8 &  20.10 &  18.89 &  18.45 &    1.42 &    1.97 &        B   \\
   241 & 1:41:55.86 & 21:32:39.8 &  20.60 &  19.66 &  19.26 &    0.92 &    1.36 &  \nodata   \\
   318 & 1:41:57.29 & 21:32:27.8 &  21.19 &  20.35 &  19.93 &    0.85 &    1.32 &        D   \\
   322 & 1:42:03.69 & 21:32:15.3 &  20.72 &  19.30 &  18.74 &    1.49 &    2.07 &        B   \\
   379 & 1:42:01.92 & 21:32:10.2 &  20.80 &  19.28 &  18.70 &    1.60 &    2.20 &        B   \\
   412 & 1:42:02.45 & 21:31:57.6 &  19.88 &  18.64 &  18.09 &    1.63 &    2.24 &        B   \\
   442 & 1:42:02.64 & 21:32:06.5 &  22.63 &  21.26 &  20.77 &    1.43 &    1.95 &        B   \\
   450 & 1:42:02.94 & 21:31:04.1 &  17.66 &  17.24 &  16.97 &    0.49 &    0.80 &  \nodata   \\
   451 & 1:42:01.13 & 21:31:55.1 &  20.62 &  19.18 &  18.63 &    1.85 &    2.54 &  \nodata   \\
   479 & 1:42:03.46 & 21:31:17.4 &  18.25 &  17.20 &  16.71 &    1.58 &    2.18 &        B   \\
   490 & 1:42:00.43 & 21:31:44.7 &  18.97 &  17.66 &  17.06 &    1.38 &    2.00 &  \nodata   \\
   537 & 1:42:08.64 & 21:31:45.5 &  21.75 &  20.42 &  19.86 &    1.44 &    2.04 &        B   \\
   614 & 1:42:01.26 & 21:31:32.0 &  21.70 &  20.21 &  19.67 &    1.56 &    2.14 &        B   \\
   637 & 1:42:01.38 & 21:31:22.1 &  21.39 &  19.88 &  19.38 &    1.56 &    2.15 &        B   \\
   671 & 1:42:03.21 & 21:31:11.9 &  19.71 &  18.29 &  17.72 &    1.66 &    2.29 &        B   \\
   760 & 1:42:01.28 & 21:31:04.6 &  21.90 &  20.41 &  19.86 &    1.56 &    2.15 &        B   \\
   777 & 1:41:59.76 & 21:30:57.9 &  21.17 &  19.64 &  19.05 &    1.65 &    2.28 &        B   \\
   844 & 1:42:07.17 & 21:30:49.7 &  21.54 &  20.14 &  19.62 &    1.49 &    2.07 &        B   \\
   900 & 1:42:09.16 & 21:30:40.6 &  21.38 &  20.78 &  20.58 &    0.78 &    1.11 &        D   \\
   911 & 1:42:03.11 & 21:30:31.6 &  22.51 &  21.12 &  20.58 &    1.43 &    1.99 &        B   \\
   971 & 1:42:06.33 & 21:30:24.9 &  21.50 &  20.86 &  20.52 &    0.70 &    1.07 &  \nodata   \\
  1012 & 1:42:01.75 & 21:30:17.4 &  22.24 &  20.73 &  20.18 &    1.53 &    2.11 &        B   \\
  1029 & 1:41:55.20 & 21:30:12.1 &  21.70 &  20.31 &  19.76 &    1.47 &    2.05 &        B   \\
  1043 & 1:41:58.57 & 21:30:01.9 &  20.70 &  19.28 &  18.74 &    1.52 &    2.11 &        B   \\
  1076 & 1:42:05.61 & 21:30:01.3 &  21.94 &  20.96 &  20.44 &    1.57 &    2.26 &  \nodata   \\
  1099 & 1:42:05.63 & 21:30:03.4 &  22.16 &  21.36 &  21.03 &    0.91 &    1.27 &        D   \\
  1154 & 1:41:54.38 & 21:29:51.1 &  21.27 &  20.38 &  20.09 &    0.95 &    1.32 &  \nodata   \\
  1179 & 1:42:00.91 & 21:29:41.6 &  20.89 &  19.85 &  19.37 &    1.16 &    1.76 &        D   \\
  1205 & 1:41:53.41 & 21:29:26.7 &  21.29 &  19.87 &  19.35 &    1.49 &    2.08 &        D   \\
  1207 & 1:42:04.04 & 21:29:35.5 &  21.85 &  20.40 &  19.82 &    1.52 &    2.13 &        B   \\
  1242 & 1:42:06.39 & 21:29:27.0 &  22.39 &  21.18 &  20.45 &    1.35 &    2.18 &  \nodata   \\
  1325 & 1:41:57.25 & 21:28:27.9 &  21.66 &  19.96 &  19.32 &    1.83 &    2.51 &  \nodata   \\
  1412 & 1:42:07.28 & 21:28:56.5 &  21.30 &  20.22 &  19.78 &    1.44 &    2.01 &        D   \\
  1416 & 1:42:06.40 & 21:28:38.7 &  21.49 &  20.33 &  19.82 &    1.31 &    2.00 &        D   \\
  1461 & 1:42:08.24 & 21:29:12.1 &  21.06 &  20.71 &  20.58 &    0.41 &    0.60 &  \nodata   \\
  1472 & 1:42:08.28 & 21:28:49.0 &  21.98 &  20.98 &  20.35 &    1.11 &    2.00 &  \nodata   \\

  \enddata

  \tablecomments{Units of right ascension are hours, minutes, and
  seconds, and units of declination are degrees, arcminutes, and
  arcseconds. Right ascension and declination are consistent with
  USNO, with an rms scatter of $\approx$ \arcsd{0}{7}. Magnitudes and
  colors are corrected for galactic extinction. B/D gives an estimate
  (for cluster members) of whether the \rfil-band morphology is
  bulge-like or disk-like (see text).}

\end{deluxetable}

Spectroscopic reductions are achieved using the Gemini IRAF package
v1.4. The method follows J05 apart from two major differences, namely
the inclusion of tilted slits and the lack of detectable
fringing. Tilted slits (\ie slits not perpendicular to the dispersion
direction) were included, aligned along the major axes of the
galaxies, for determination of galactic rotation curves. An analysis
of this aspect of the data, as well as a determination of a
Tully-Fisher relation for \rxj \ will be presented in a future
paper. At the present time we merely outline the extra steps in the
reduction process which result from this intricacy.

Bias-subtraction and overscan-trim are carried out in standard fashion
using the same method as the imaging data. Flatfields are created from
lamp spectra and applied on a chip-by-chip basis.

To subtract the sky, which includes strong emission lines, we use a
technique which avoids the interpolation of the data which occurs
during wavelength calibration. A second-order Chebyshev polynomial is
fit in the spatial direction column by column, rejecting points $\pm 3
\sigma$ above and below the fit and an object-centred aperture of
between \arcsd{1}{4} and \arcsd{6}{1} depending on the spatial extent
of the object. This sky is then subtracted.

For objects observed through tilted slits the spectra must be
rectified in order to fit the sky. The angle of the slit on the mask
and the angle of lines in the dispersed spectrum differ by the
anamorphic factor, which is a function of grating angle and slit
position. As a consequence, we use between 3 and 5 strong skylines in
each spectrum to determine the best rectification angle on a
spectrum-by-spectrum basis. This is typically between 80\% and 90\% of
the tilt in the slit mask. Spectra are straightened by shifting
individual rows in multiples of a fifth of a pixel. The sky is then
subtracted as described above, and the spectra are restored to their
original orientation.

Subtracted sky spectra are retained and processed in the same way
as the science from this point. 

The sky-subtracted exposures taken at the same central wavelength
setting are combined. Bad pixels and cosmic rays are rejected at this
stage. Then the images are mosaicked using the correct transformation
for the relative position of the GMOS-N CCDs. The spectra are
wavelength-calibrated using CuAr spectra taken using the same
instrumental setup as the science observations. Calibrated spectra at
each central wavelength setting are co-added and those objects which
appear in both masks are also combined.

The 2D spectra are traced and extracted as 1D spectra using an
aperture of \arcsd{1}{2} centred on the maximum signal perpendicular
to the spectrum.  Atmospheric telluric absorption lines are corrected
for by combining all spectra into a single spectrum. A 27-piece cubic
spline fit is used to normalise the resulting spectrum with pixels
$\pm 3 \sigma$ from the fit rejected. Regions unaffected by telluric
lines are set to unity. Figure~\ref{fig:tel} shows the combined
spectrum and the function used for telluric correction.

\begin{figure*}

  \begin{center}

    \epsscale{1}
    \plotone{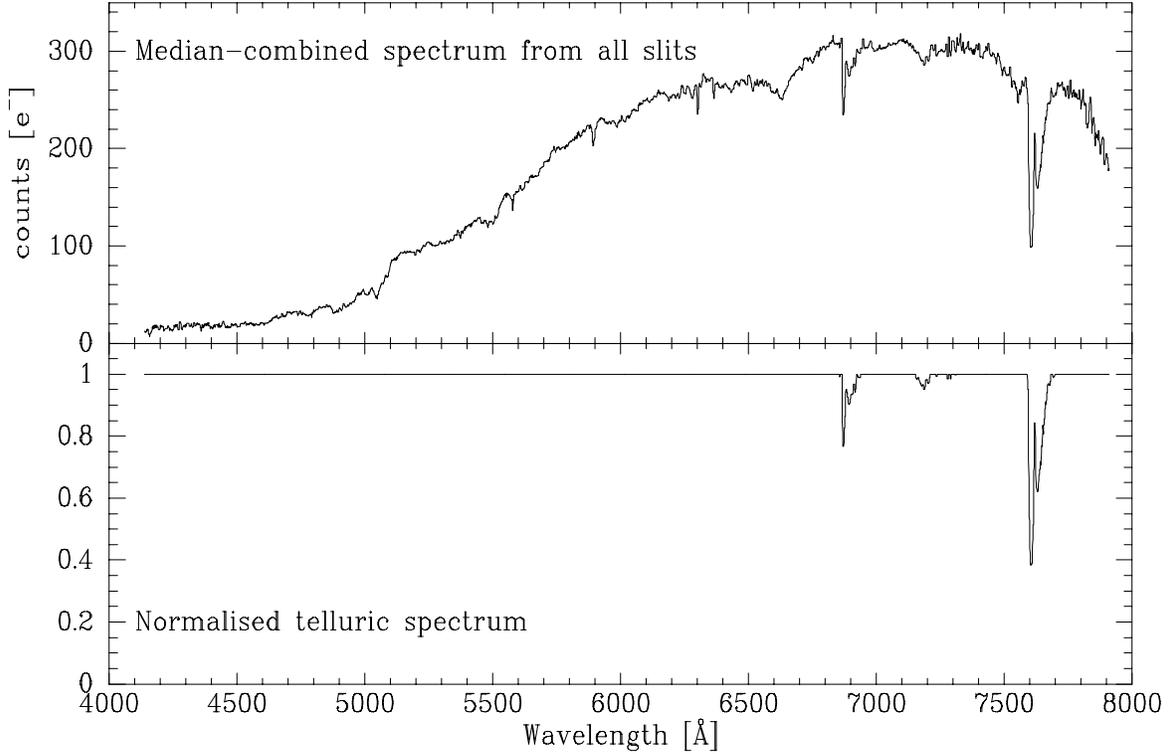}

    \figcaption{{\em Top:} Spectrum formed from median-combining spectra
    of all objects. {\em Bottom:} Normalised telluric absorption
    spectrum.\label{fig:tel}}

   \end{center}

\end{figure*}

The instrumental resolution for each slit ranges from 1.4 -- 1.6\AA,
measured from Gaussian fits to the sky lines. Spectra are resampled to
0.92\AA \ per pixel to improve signal-to-noise. A relative flux
calibration is then achieved using a sensitivity function covering the
full spectral range of the MOS spectra (see J05 for details).

\section{GMOS-N spectroscopic results}

Derived redshifts, velocity dispersions and stellar template fractions
for all galaxies are given in Table~\ref{tab:kin}. Table~\ref{tab:lin}
gives the line indices and corresponding uncertainties for the
spectroscopic sample, excluding the M star.

Figure~\ref{fig:sp1} show rest-frame object and noise spectra for the
30 cluster members in our spectroscopic sample together with their
magnitudes, colors and velocity dispersions. Figure~\ref{fig:sp2} is
the corresponding figure for non-members. The spectrum of the M star
is not shown.

Postage stamp images of the cluster members in the spectroscopic
sample are presented in Figure~\ref{fig:crm}. These are constructed
from the GMOS-N \gfil, \rfil, and \ifil \ filters. The equivalent
postage stamps for non-members are shown in Figure~\ref{fig:crn}.

\begin{deluxetable}{rcccccccc}

  \tableheadfrac{0}
  \tablewidth{0pt}
  \tablecolumns{9} 
  \tablecaption{Results from the template fitting\label{tab:kin}}

  \tablehead{\colhead{ID} & \colhead{$z$} &
   \colhead{Member$^{\mathrm{a}}$} & \colhead{log $\sigma$} &
   \colhead{log $\sigma_{\mathrm{corr}}^{\mathrm{b}}$} &
   \colhead{$\sigma_{\mathrm{log }\sigma}$} &
   \multicolumn{3}{c}{Template fractions} \\ \colhead{} & \colhead{} &
   \colhead{} & \colhead{} & \colhead{} & \colhead{} & \colhead{B8V} &
   \colhead{G1V} & \colhead{K0III}}

  \startdata

         1 &   0.2823 &     1 &   2.291 &   2.309 &   0.016 &   0.00 &   0.68 &   0.32   \\
    22 &   0.2816 &     1 &   2.290 &   2.308 &   0.020 &   0.00 &   0.89 &   0.11   \\
    44 &   0.4996 &     0 &   2.048 &   2.072 &   0.073 &   0.00 &   0.00 &   1.00   \\
    88 &   0.2846 &     1 &   1.879 &   1.897 &   0.116 &   0.41 &   0.59 &   0.00   \\
   116 &   0.2730 &     1 &   1.939 &   1.957 &   0.075 &   0.34 &   0.12 &   0.53   \\
   128 &   0.2808 &     1 &   2.210 &   2.228 &   0.023 &   0.00 &   0.66 &   0.34   \\
   173 &   0.2991 &     0 &   2.120 &   2.139 &   0.077 &   0.00 &   0.78 &   0.22   \\
   205 &   0.2736 &     1 &   2.156 &   2.173 &   0.021 &   0.00 &   0.60 &   0.40   \\
   241 &   0.2659 &     0 &   1.797 &   1.814 &   0.055 &   0.44 &   0.46 &   0.10   \\
   318 &   0.2823 &     1 &   2.112 &   2.130 &   0.030 &   0.49 &   0.44 &   0.07   \\
   322 &   0.2825 &     1 &   1.967 &   1.985 &   0.025 &   0.00 &   0.67 &   0.33   \\
   379 &   0.2857 &     1 &   2.238 &   2.256 &   0.017 &   0.00 &   0.49 &   0.51   \\
   412 &   0.2797 &     1 &   2.238 &   2.256 &   0.019 &   0.00 &   0.52 &   0.48   \\
   442 &   0.2786 &     1 &   1.795 &   1.813 &   0.166 &   0.00 &   1.00 &   0.00   \\
   450 &   0.0696 &     0 &   2.078 &   2.077 &   0.055 &   0.47 &   0.53 &   0.00   \\
   451 &   0.3853 &     0 &   2.225 &   2.246 &   0.016 &   0.00 &   0.53 &   0.47   \\
   479 &   0.2750 &     1 &   2.154 &   2.171 &   0.023 &   0.00 &   0.53 &   0.47   \\
   490 &   0.0000 &     0 &  \nodata &  \nodata &  \nodata &  \nodata &  \nodata &  \nodata   \\
   537 &   0.2784 &     1 &   2.140 &   2.157 &   0.026 &   0.00 &   0.90 &   0.10   \\
   614 &   0.2727 &     1 &   2.126 &   2.144 &   0.019 &   0.00 &   0.67 &   0.33   \\
   637 &   0.2782 &     1 &   1.957 &   1.975 &   0.037 &   0.00 &   0.50 &   0.50   \\
   671 &   0.2839 &     1 &   2.344 &   2.362 &   0.018 &   0.00 &   0.34 &   0.66   \\
   760 &   0.2800 &     1 &   2.169 &   2.187 &   0.020 &   0.00 &   0.67 &   0.33   \\
   777 &   0.2823 &     1 &   2.317 &   2.335 &   0.023 &   0.00 &   0.62 &   0.38   \\
   844 &   0.2712 &     1 &   2.157 &   2.175 &   0.020 &   0.00 &   0.80 &   0.20   \\
   900 &   0.2886 &     1 &   1.961 &   1.979 &   0.052 &   0.54 &   0.46 &   0.00   \\
   911 &   0.2775 &     1 &   1.740 &   1.758 &   0.215 &   0.00 &   0.73 &   0.27   \\
   971 &   0.6151 &     0 &   2.502 &   2.527 &   0.067 &   0.72 &   0.28 &   0.00   \\
  1012 &   0.2762 &     1 &   1.931 &   1.949 &   0.037 &   0.00 &   0.83 &   0.17   \\
  1029 &   0.2834 &     1 &   1.806 &   1.824 &   0.035 &   0.00 &   0.53 &   0.47   \\
  1043 &   0.2715 &     1 &   2.037 &   2.055 &   0.025 &   0.00 &   0.50 &   0.50   \\
  1076 &   0.4036 &     0 &   1.910 &   1.932 &   0.055 &   0.00 &   0.83 &   0.17   \\
  1099 &   0.2895 &     1 &   2.117 &   2.135 &   0.153 &   0.47 &   0.36 &   0.18   \\
  1154 &   0.3869 &     0 &   2.149 &   2.170 &   0.058 &   0.56 &   0.41 &   0.03   \\
  1179 &   0.2825 &     1 &   1.803 &   1.821 &   0.068 &   0.10 &   0.37 &   0.52   \\
  1205 &   0.2706 &     1 &   1.861 &   1.879 &   0.042 &   0.00 &   0.65 &   0.35   \\
  1207 &   0.2767 &     1 &   2.092 &   2.110 &   0.035 &   0.00 &   0.24 &   0.76   \\
  1242 &   0.6155 &     0 &   2.510 &   2.534 &   0.057 &   0.50 &   0.50 &   0.00   \\
  1325 &   0.3875 &     0 &   2.280 &   2.302 &   0.042 &   0.00 &   0.66 &   0.34   \\
  1412 &   0.2728 &     1 &   1.973 &   1.991 &   0.043 &   0.00 &   1.00 &   0.00   \\
  1416 &   0.2864 &     1 &   2.004 &   2.022 &   0.103 &   0.00 &   0.00 &   1.00   \\
  1461 &   0.1896 &     0 &   2.622 &   2.635 &   0.085 &   0.24 &   0.76 &   0.00   \\
  1472 &   0.6879 &     0 &   2.478 &   2.504 &   0.076 &   0.00 &   0.66 &   0.34   \\

  \enddata
 
  \tablecomments{$^{\mathrm{a}}$ 1 -- Galaxy is a member of \rxj, 0 --
  spectroscopic target is not a member of \rxj. \\ $^{\mathrm{b}}$
  Velocity dispersion corrected to a standard-sized aperture
  equivalent to a circular aperture with a diameter of \arcsd{3}{4} at
  the distance of the Coma cluster.}

\end{deluxetable}

\LongTables

\begin{deluxetable}{rrrrrrrrrrrrrr}
  \tabletypesize{\footnotesize}
  \tablewidth{0pt}
  \tablecolumns{14}
  \tablecaption{Line indices\label{tab:lin}}
  \tablehead{\colhead{ID} & \colhead{CN3883} & \colhead{CaHK} &
    \colhead{D4000} & \colhead{H$\delta_{\mathrm{A}}$} &
    \colhead{\CNtwo} & \colhead{G4300} &
    \colhead{H$\gamma_{\mathrm{A}}$} & \colhead{C4668} &
    \colhead{\Hbetaem} & \colhead{\mgb} & \colhead{Fe5270} &
    \colhead{Fe5335} & \colhead{\fe}}

  \startdata

                          1$\: \: \: \: \: $ &  \nodata & \nodata & \nodata &  -2.870 &   0.140 &   5.810 &  -5.660 &   8.300 &   1.810 &   4.970 &   2.730 &   1.970 &   2.351   \\*
                      1$\: \: \: \: \: $ &  \nodata & \nodata & \nodata &   0.159 &   0.005 &   0.120 &   0.238 &   0.131 &   0.062 &   0.073 &   0.074 &   0.075 &   0.053   \\
                     22$\: \: \: \: \: $ &  \nodata &  20.790 & \nodata &  -1.640 &   0.190 &   5.820 &  -6.180 &   4.340 &   1.420 &   5.020 &   2.940 &   1.550 &   2.242   \\*
                     22$\: \: \: \: \: $ &  \nodata &   0.404 & \nodata &   0.217 &   0.006 &   0.154 &   0.323 &   0.152 &   0.069 &   0.079 &   0.077 &   0.078 &   0.055   \\
                     44$\: \: \: \: \: $ &    0.070 &  -1.740 &   1.180 &   2.970 & \nodata &   0.610 &  -3.770 &   3.490 &  -0.820 &   1.290 & \nodata & \nodata & \nodata   \\*
                     44$\: \: \: \: \: $ &    0.010 &   0.969 &   0.009 &   0.601 & \nodata &   0.491 &   0.560 &   0.707 &   0.452 &   0.624 & \nodata & \nodata & \nodata   \\
                     88$\: \: \: \: \: $ &  \nodata & \nodata & \nodata &   9.360 & \nodata &   0.500 &  -1.810 &   5.800 &   2.050 &   1.660 &   4.100 &   2.170 &   3.137   \\*
                     88$\: \: \: \: \: $ &  \nodata & \nodata & \nodata &   0.911 & \nodata &   0.665 &   1.474 &   1.194 &   0.492 &   0.682 &   0.702 &   0.797 &   0.531   \\
                    116$^{\mathrm{a,b}}$ &  \nodata &  11.790 &   1.630 &  -1.680 &   0.020 &   1.350 &   0.050 &  -5.600 &   0.730 &   3.310 &   0.240 & \nodata &   0.222   \\*
                    116$\: \: \: \: \: $ &  \nodata &   1.468 &   0.017 &   1.061 &   0.027 &   0.808 &   0.877 &   1.093 &   0.427 &   0.493 &   0.550 & \nodata &   0.506   \\
            128$^{\mathrm{a}} \: \: \: $ &  \nodata &  19.980 & \nodata &  -0.550 &   0.070 &   6.140 &  -4.710 &   6.080 &   1.730 &   4.630 &   2.530 & \nodata &   2.326   \\*
                    128$\: \: \: \: \: $ &  \nodata &   0.531 & \nodata &   0.295 &   0.009 &   0.228 &   0.523 &   0.243 &   0.113 &   0.129 &   0.129 & \nodata &   0.118   \\
            173$^{\mathrm{b}} \: \: \: $ &    0.120 &  13.880 &   1.830 &   3.520 &   0.020 &   2.400 &  -0.870 &   5.200 &  -0.290 &   1.810 &   2.770 & \nodata & \nodata   \\*
                    173$\: \: \: \: \: $ &    0.010 &   0.585 &   0.007 &   0.338 &   0.010 &   0.539 &   0.304 &   0.313 &   0.201 &   0.171 &   0.208 & \nodata & \nodata   \\
                    205$\: \: \: \: \: $ &    0.220 &  20.880 &   2.030 &  -0.770 &   0.100 &   5.170 &  -4.020 &   5.070 &   2.370 &   2.860 &   2.700 &   3.030 &   2.866   \\*
                    205$\: \: \: \: \: $ &    0.006 &   0.328 &   0.004 &   0.197 &   0.006 &   0.150 &   0.180 &   0.175 &   0.077 &   0.103 &   0.100 &   0.094 &   0.069   \\
            241$^{\mathrm{b}} \: \: \: $ &    0.090 &   9.530 &   1.410 &   3.410 & \nodata &   0.950 &   2.720 &   2.360 &   0.530 &   2.240 &   2.070 &   1.900 &   1.982   \\*
                    241$\: \: \: \: \: $ &    0.006 &   0.415 &   0.004 &   0.263 & \nodata &   0.223 &   0.235 &   0.282 &   0.129 &   0.153 &   0.152 &   0.145 &   0.105   \\
            318$^{\mathrm{b}} \: \: \: $ &    0.040 &   9.450 &   1.330 &   3.320 & \nodata &   2.430 &  -0.070 &   3.040 &  -4.790 &   2.830 &   1.490 &   0.490 &   0.991   \\*
                    318$\: \: \: \: \: $ &    0.004 &   0.306 &   0.003 &   0.203 & \nodata &   0.179 &   0.360 &   0.219 &   0.118 &   0.122 &   0.131 &   0.138 &   0.095   \\
            322$^{\mathrm{c}} \: \: \: $ &    0.250 &  23.050 &   2.280 &   0.550 &   0.080 &  -2.470 &   5.610 &   9.260 &   2.000 &   4.480 &   3.320 &   1.850 &   2.586   \\*
                    322$\: \: \: \: \: $ &    0.014 &   0.674 &   0.009 &   0.403 &   0.011 &   0.321 &   0.572 &   0.284 &   0.137 &   0.163 &   0.158 &   0.165 &   0.114   \\
                    379$\: \: \: \: \: $ &    0.270 &  23.260 &   2.320 &  -0.940 &   0.090 &   5.030 &  -5.020 &   7.380 &   1.990 &   4.610 &   3.190 &   2.800 &   2.998   \\*
                    379$\: \: \: \: \: $ &    0.006 &   0.305 &   0.004 &   0.176 &   0.005 &   0.135 &   0.264 &   0.132 &   0.063 &   0.077 &   0.075 &   0.075 &   0.053   \\
                    412$\: \: \: \: \: $ &    0.290 &  22.510 &   2.260 &  -1.540 &   0.140 &   6.290 &  -6.330 &   7.800 &   1.590 &   4.900 &   2.870 &   2.440 &   2.651   \\*
                    412$\: \: \: \: \: $ &    0.006 &   0.269 &   0.004 &   0.164 &   0.005 &   0.123 &   0.244 &   0.130 &   0.060 &   0.076 &   0.075 &   0.075 &   0.053   \\
                    442$\: \: \: \: \: $ &    0.250 &  17.760 &   2.670 &  -0.750 &   0.030 &   9.040 & -10.190 &   3.840 &   2.500 &   2.420 &   1.250 &   2.190 &   1.721   \\*
                    442$\: \: \: \: \: $ &    0.037 &   1.851 &   0.024 &   0.908 &   0.025 &   0.591 &   1.389 &   0.684 &   0.297 &   0.389 &   0.367 &   0.384 &   0.266   \\
            450$^{\mathrm{b}} \: \: \: $ &  \nodata & \nodata & \nodata & \nodata & \nodata & \nodata & \nodata &   1.280 &  -2.350 &   3.270 &   2.090 &   1.260 &   1.673   \\*
                    450$\: \: \: \: \: $ &  \nodata & \nodata & \nodata & \nodata & \nodata & \nodata & \nodata &   0.219 &   0.110 &   0.115 &   0.128 &   0.128 &   0.091   \\
                    451$\: \: \: \: \: $ &    0.230 &  24.620 &   2.280 &  -0.560 &   0.070 &   5.230 &  -5.050 &   9.460 &   2.260 &   4.250 & \nodata & \nodata & \nodata   \\*
                    451$\: \: \: \: \: $ &    0.007 &   0.372 &   0.005 &   0.212 &   0.006 &   0.170 &   0.193 &   0.172 &   0.079 &   0.104 & \nodata & \nodata & \nodata   \\
                    479$\: \: \: \: \: $ &    0.300 &  21.140 &   2.190 &  -1.290 &   0.130 &   4.990 &  -4.470 &   6.510 &   1.710 &   5.100 &   2.500 &   2.940 &   2.719   \\*
                    479$\: \: \: \: \: $ &    0.006 &   0.315 &   0.004 &   0.189 &   0.006 &   0.143 &   0.169 &   0.156 &   0.071 &   0.091 &   0.091 &   0.084 &   0.062   \\
                    537$\: \: \: \: \: $ &  \nodata & \nodata & \nodata &  -1.460 &   0.100 &   5.370 &  -4.460 &   2.300 &   1.750 &   4.140 &   2.590 &   2.280 &   2.437   \\*
                    537$\: \: \: \: \: $ &  \nodata & \nodata & \nodata &   0.373 &   0.011 &   0.285 &   0.586 &   0.296 &   0.130 &   0.155 &   0.151 &   0.152 &   0.107   \\
                    614$\: \: \: \: \: $ &    0.250 &  25.110 &   2.390 &  -2.130 &   0.110 &   5.040 &  -3.800 &   5.660 &   2.190 &   4.640 &   2.520 &   2.700 &   2.610   \\*
                    614$\: \: \: \: \: $ &    0.008 &   0.383 &   0.005 &   0.220 &   0.006 &   0.165 &   0.194 &   0.165 &   0.070 &   0.093 &   0.089 &   0.082 &   0.061   \\
                    637$\: \: \: \: \: $ &    0.290 &  20.990 &   2.130 &  -1.560 &   0.110 &   5.810 &  -5.100 &   6.360 &   1.700 &   4.490 &   2.430 &   2.430 &   2.432   \\*
                    637$\: \: \: \: \: $ &    0.010 &   0.507 &   0.007 &   0.294 &   0.008 &   0.222 &   0.404 &   0.220 &   0.097 &   0.125 &   0.122 &   0.119 &   0.085   \\
            671$^{\mathrm{c}} \: \: \: $ &    0.290 &  22.810 &   2.520 &  -1.330 &   0.130 &   5.860 &  -4.240 &   8.480 &   1.810 &   4.660 &   2.990 &   2.460 &   2.724   \\*
                    671$\: \: \: \: \: $ &    0.004 &   0.177 &   0.003 &   0.104 &   0.003 &   0.077 &   0.132 &   0.083 &   0.040 &   0.047 &   0.047 &   0.047 &   0.034   \\
                    760$\: \: \: \: \: $ &    0.180 &  24.710 &   2.680 &  -2.060 &   0.110 &   8.150 &  -5.460 &   8.480 &   2.070 &   5.100 &   3.270 &   1.950 &   2.609   \\*
                    760$\: \: \: \: \: $ &    0.016 &   0.704 &   0.011 &   0.379 &   0.011 &   0.284 &   0.621 &   0.264 &   0.119 &   0.150 &   0.142 &   0.143 &   0.101   \\
                    777$\: \: \: \: \: $ &    0.210 &  20.860 &   2.310 &  -1.930 &   0.130 &   6.600 &  -6.110 &   8.580 &   3.050 &   5.180 &   3.850 &   2.010 &   2.931   \\*
                    777$\: \: \: \: \: $ &    0.009 &   0.452 &   0.006 &   0.245 &   0.008 &   0.184 &   0.372 &   0.171 &   0.076 &   0.094 &   0.089 &   0.094 &   0.065   \\
                    844$\: \: \: \: \: $ &  \nodata &  23.290 & \nodata &  -1.240 &   0.110 &   4.790 &  -4.700 &   6.130 &   2.310 &   3.930 &   2.640 &   2.820 &   2.730   \\*
                    844$\: \: \: \: \: $ &  \nodata &   0.380 & \nodata &   0.225 &   0.006 &   0.166 &   0.195 &   0.183 &   0.076 &   0.098 &   0.094 &   0.089 &   0.065   \\
            900$^{\mathrm{b}} \: \: \: $ &  \nodata & \nodata & \nodata &   4.400 & \nodata &   0.650 &   2.180 &  -1.340 &  -2.200 &   2.580 &   2.160 &   3.500 &   2.829   \\*
                    900$\: \: \: \: \: $ &  \nodata & \nodata & \nodata &   0.484 & \nodata &   0.474 &   1.004 &   0.635 &   0.315 &   0.353 &   0.365 &   0.399 &   0.270   \\
                    911$\: \: \: \: \: $ &    0.680 &  24.200 &   4.730 &  -1.800 & \nodata &   5.030 &  -3.810 &   2.490 &   1.820 &   4.370 &   2.260 &   2.130 &   2.195   \\*
                    911$\: \: \: \: \: $ &    0.039 &   0.888 &   0.021 &   0.438 & \nodata &   0.310 &   0.388 &   0.358 &   0.159 &   0.177 &   0.177 &   0.184 &   0.128   \\
                    971$\: \: \: \: \: $ &  \nodata &   6.100 &   1.240 &  -0.030 &   0.040 &   2.100 &  -6.620 & \nodata & \nodata & \nodata & \nodata & \nodata & \nodata   \\*
                    971$\: \: \: \: \: $ &  \nodata &   0.353 &   0.003 &   0.215 &   0.006 &   0.199 &   0.228 & \nodata & \nodata & \nodata & \nodata & \nodata & \nodata   \\
                   1012$\: \: \: \: \: $ &    0.410 &  26.400 &   2.740 &  -0.770 &   0.050 &   4.020 &  -2.420 &  -0.030 &   2.450 &   3.870 &   2.940 &   2.620 &   2.777   \\*
                   1012$\: \: \: \: \: $ &    0.014 &   0.541 &   0.008 &   0.282 &   0.008 &   0.256 &   0.288 &   0.287 &   0.117 &   0.137 &   0.132 &   0.126 &   0.091   \\
           1029$^{\mathrm{a}} \: \: \: $ &    0.120 &  22.070 &   1.730 &  -1.380 &   0.120 &   5.840 &  -5.330 &   4.030 &   2.110 &   3.440 &   2.930 & \nodata &   2.691   \\*
                   1029$\: \: \: \: \: $ &    0.008 &   0.532 &   0.006 &   0.312 &   0.009 &   0.233 &   0.493 &   0.225 &   0.105 &   0.121 &   0.118 & \nodata &   0.108   \\
                   1043$\: \: \: \: \: $ &    0.250 &  22.120 &   2.050 &  -0.420 &   0.110 &   4.870 &  -4.170 &   7.040 &   2.410 &   4.600 &   3.070 &   2.680 &   2.876   \\*
                   1043$\: \: \: \: \: $ &    0.006 &   0.320 &   0.004 &   0.203 &   0.006 &   0.152 &   0.175 &   0.158 &   0.069 &   0.087 &   0.086 &   0.082 &   0.059   \\
                   1076$^{\mathrm{a,b}}$ &    0.200 &  16.770 &   2.250 &  -2.390 &   0.120 &   1.370 &  -0.370 &  -2.710 &   0.370 &   3.780 &   3.200 & \nodata &   2.939   \\*
                   1076$\: \: \: \: \: $ &    0.021 &   1.709 &   0.016 &   0.668 &   0.019 &   0.518 &   0.543 &   0.606 &   0.306 &   0.451 &   0.416 & \nodata &   0.383   \\
           1099$^{\mathrm{b}} \: \: \: $ &    0.180 &  11.350 &   1.470 &   4.460 & \nodata &   2.760 &  -3.510 &   7.540 &  -9.030 &   9.470 &   2.810 &   3.200 &   3.007   \\*
                   1099$\: \: \: \: \: $ &    0.028 &   1.676 &   0.017 &   1.008 & \nodata &   1.028 &   2.362 &   1.318 &   0.764 &   0.860 &   1.137 &   1.276 &   0.854   \\
           1154$^{\mathrm{b}} \: \: \: $ &  \nodata &   7.210 &   1.520 &   3.000 & \nodata &   0.920 &   0.840 &   1.970 & \nodata & \nodata & \nodata & \nodata & \nodata   \\*
                   1154$\: \: \: \: \: $ &  \nodata &   0.382 &   0.004 &   0.226 & \nodata &   0.217 &   0.234 &   0.268 & \nodata & \nodata & \nodata & \nodata & \nodata   \\
                   1179$^{\mathrm{b,c}}$ &    0.140 &  13.760 &   1.800 &  -1.110 & \nodata &   4.360 &  -8.570 &   5.200 &  -3.730 &   3.580 &   2.720 &   1.920 &   2.324   \\*
                   1179$\: \: \: \: \: $ &    0.016 &   0.974 &   0.011 &   0.550 & \nodata &   0.538 &   1.176 &   0.625 &   0.308 &   0.273 &   0.274 &   0.290 &   0.200   \\
                   1205$\: \: \: \: \: $ &    0.160 &  23.220 &   2.080 &  -1.380 &   0.150 &   5.260 &  -4.270 &   9.080 &   2.720 &   4.050 & \nodata & \nodata & \nodata   \\*
                   1205$\: \: \: \: \: $ &    0.011 &   0.573 &   0.007 &   0.335 &   0.009 &   0.230 &   0.270 &   0.243 &   0.098 &   0.123 & \nodata & \nodata & \nodata   \\
                   1207$\: \: \: \: \: $ &    0.250 &  26.120 &   2.530 &  -0.810 &   0.100 &   5.910 &  -7.060 &   5.810 &   1.230 &   1.840 &   2.670 &   2.300 &   2.486   \\*
                   1207$\: \: \: \: \: $ &    0.011 &   0.480 &   0.007 &   0.275 &   0.008 &   0.212 &   0.261 &   0.227 &   0.100 &   0.129 &   0.119 &   0.116 &   0.083   \\
                   1242$\: \: \: \: \: $ &    0.080 &  10.100 &   1.600 &   5.370 & \nodata &   9.000 &  -0.800 & \nodata & \nodata & \nodata & \nodata & \nodata & \nodata   \\*
                   1242$\: \: \: \: \: $ &    0.012 &   0.819 &   0.008 &   0.441 & \nodata &   0.427 &   0.433 & \nodata & \nodata & \nodata & \nodata & \nodata & \nodata   \\
           1325$^{\mathrm{b}} \: \: \: $ &    0.160 &  20.270 &   2.100 &  -1.700 &   0.130 &   4.910 &  -2.030 &   8.770 &   2.420 & \nodata & \nodata & \nodata & \nodata   \\*
                   1325$\: \: \: \: \: $ &    0.005 &   0.295 &   0.004 &   0.166 &   0.005 &   0.130 &   0.144 &   0.132 &   0.060 & \nodata & \nodata & \nodata & \nodata   \\
           1412$^{\mathrm{c}} \: \: \: $ &  \nodata &  21.970 & \nodata &  -1.020 &   0.100 &   4.970 &  -2.180 &   4.510 &   2.750 &   3.700 &   2.620 &   3.890 &   3.253   \\*
                   1412$\: \: \: \: \: $ &  \nodata &   0.434 & \nodata &   0.288 &   0.008 &   0.200 &   0.237 &   0.227 &   0.096 &   0.125 &   0.116 &   0.106 &   0.079   \\
           1416$^{\mathrm{b}} \: \: \: $ &    0.310 &  33.550 &   2.660 &   3.050 & \nodata &   3.380 &  -0.140 &   9.430 &  -0.780 &   3.620 &  -1.220 &   2.850 &   0.818   \\*
                   1416$\: \: \: \: \: $ &    0.023 &   1.021 &   0.014 &   0.518 & \nodata &   0.346 &   0.751 &   0.363 &   0.174 &   0.186 &   0.186 &   0.192 &   0.134   \\
           1461$^{\mathrm{b}} \: \: \: $ &  \nodata & \nodata & \nodata & \nodata & \nodata &   2.590 &  -1.470 &   4.360 &  -7.770 &  -1.790 &   1.700 &   3.110 &   2.406   \\*
                   1461$\: \: \: \: \: $ &  \nodata & \nodata & \nodata & \nodata & \nodata &   0.262 &   0.288 &   0.503 &   0.161 &   0.209 &   0.232 &   0.284 &   0.183   \\
                   1472$\: \: \: \: \: $ &    0.120 &  16.390 &   1.600 &   6.540 &   0.050 &   4.640 &  -2.820 & \nodata & \nodata & \nodata & \nodata & \nodata & \nodata   \\*
                   1472$\: \: \: \: \: $ &    0.005 &   0.320 &   0.004 &   0.220 &   0.006 &   0.245 &   0.273 & \nodata & \nodata & \nodata & \nodata & \nodata & \nodata   \\

  \enddata

  \tablecomments{For each object, the first line gives the value of the
  line index and the second line denotes the uncertainties. \\ $^{a}$
  \fe \ is calculated from Fe5270 only assuming the best fitting linear
  relation between Fe5270 and Fe5335 derived from cluster
  members. $^{\mathrm{b}}$ Emission-line galaxy. $^{\mathrm{c}}$ \HdgA
  \ is derived from \HdeltaA \ only assuming the best fitting linear
  relation between \HdeltaA \ and \HdgA \ for cluster members.}

\end{deluxetable}

\begin{figure}

  \begin{center}

    \epsscale{1}
    \plotone{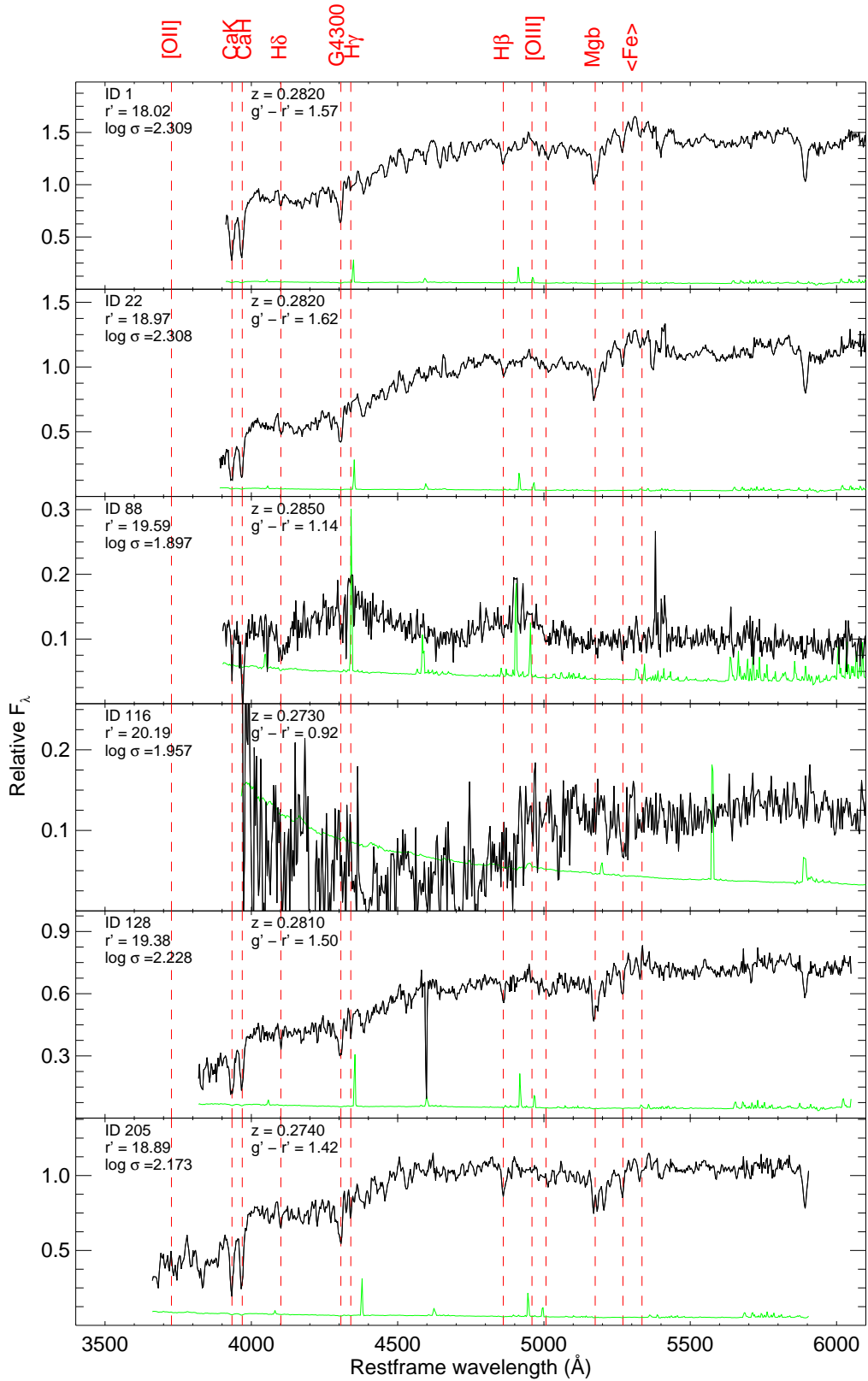}

    \figcaption{Extracted 1D spectra of cluster members in the rest
    frame. The black line represents the flux-calibrated spectrum, the
    green line is noise multiplied by a factor of two for
    clarity. The positions of the most prominent spectral features are
    labeled with vertical dashed lines.\label{fig:sp1}}

  \end{center}

\end{figure}

\setcounter{figure}{9}
\begin{figure}

  \begin{center}

    \plotone{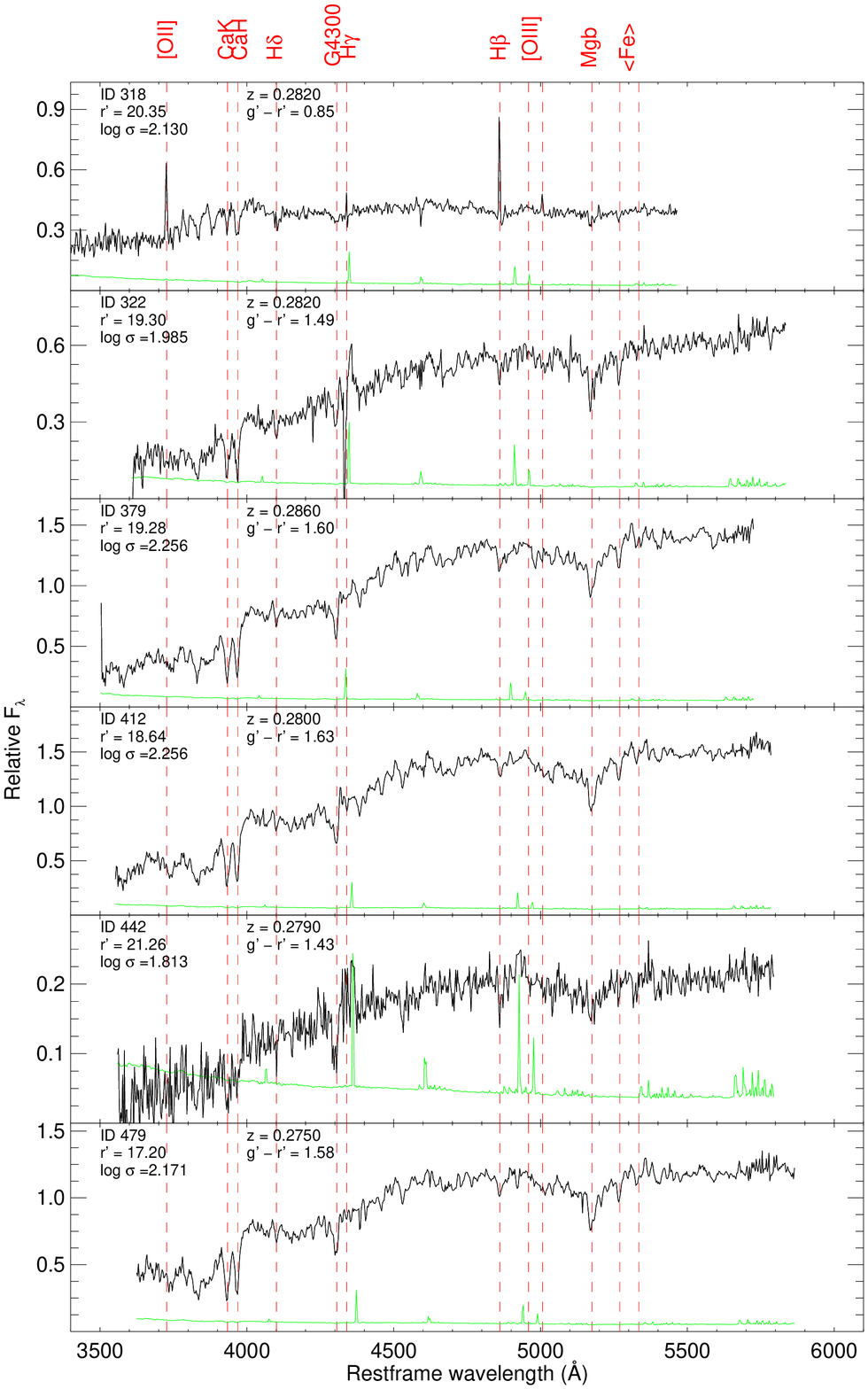}
    \figcaption{Continued.}

  \end{center}

\end{figure}

\setcounter{figure}{9}
\begin{figure}

  \begin{center}

    \plotone{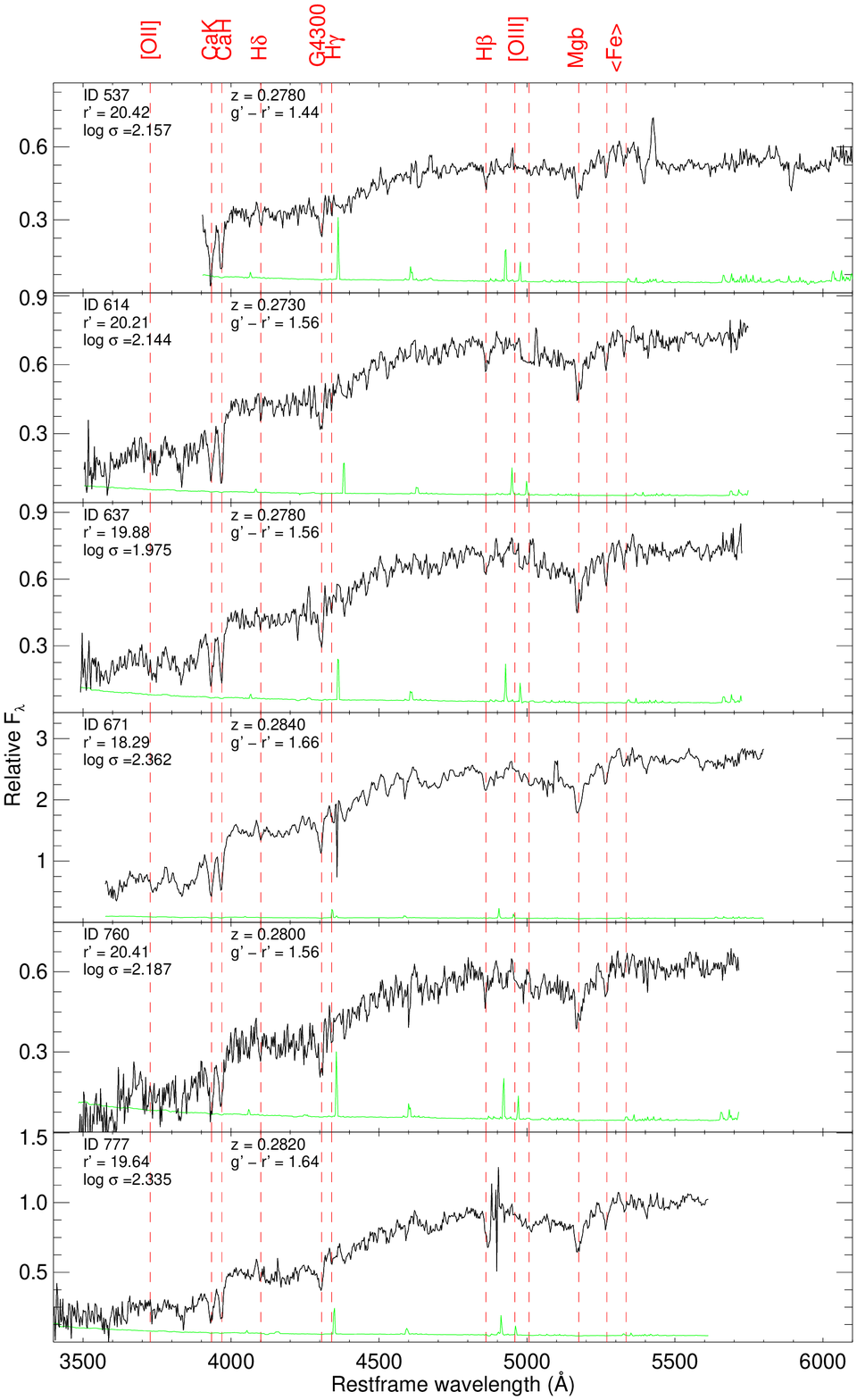}
    \figcaption{Continued.}

  \end{center}

\end{figure}

\setcounter{figure}{9}
\begin{figure}

  \begin{center}

    \plotone{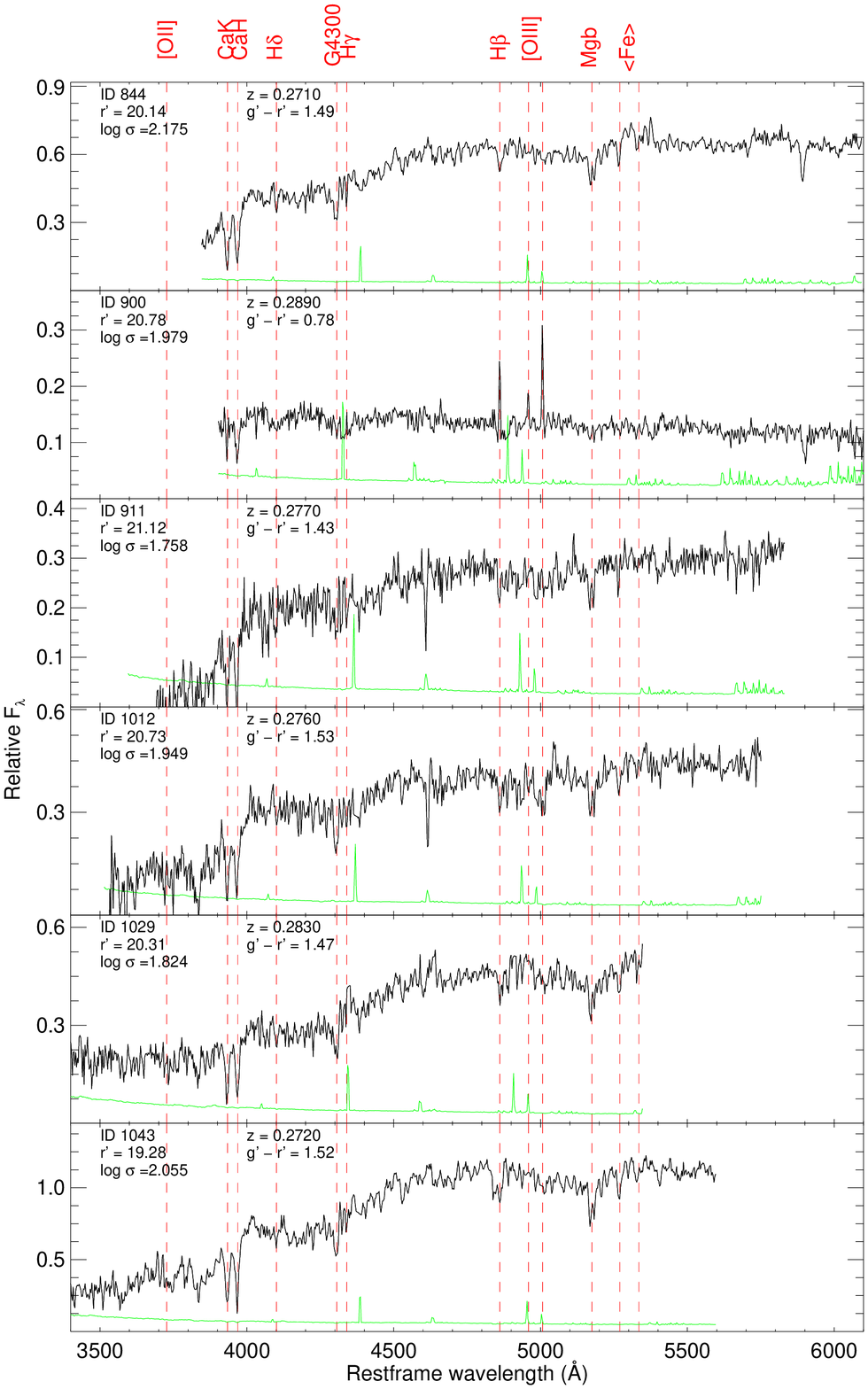}
    \figcaption{Continued.}

 \end{center}

\end{figure}

\setcounter{figure}{9}
\begin{figure}

  \begin{center}

    \plotone{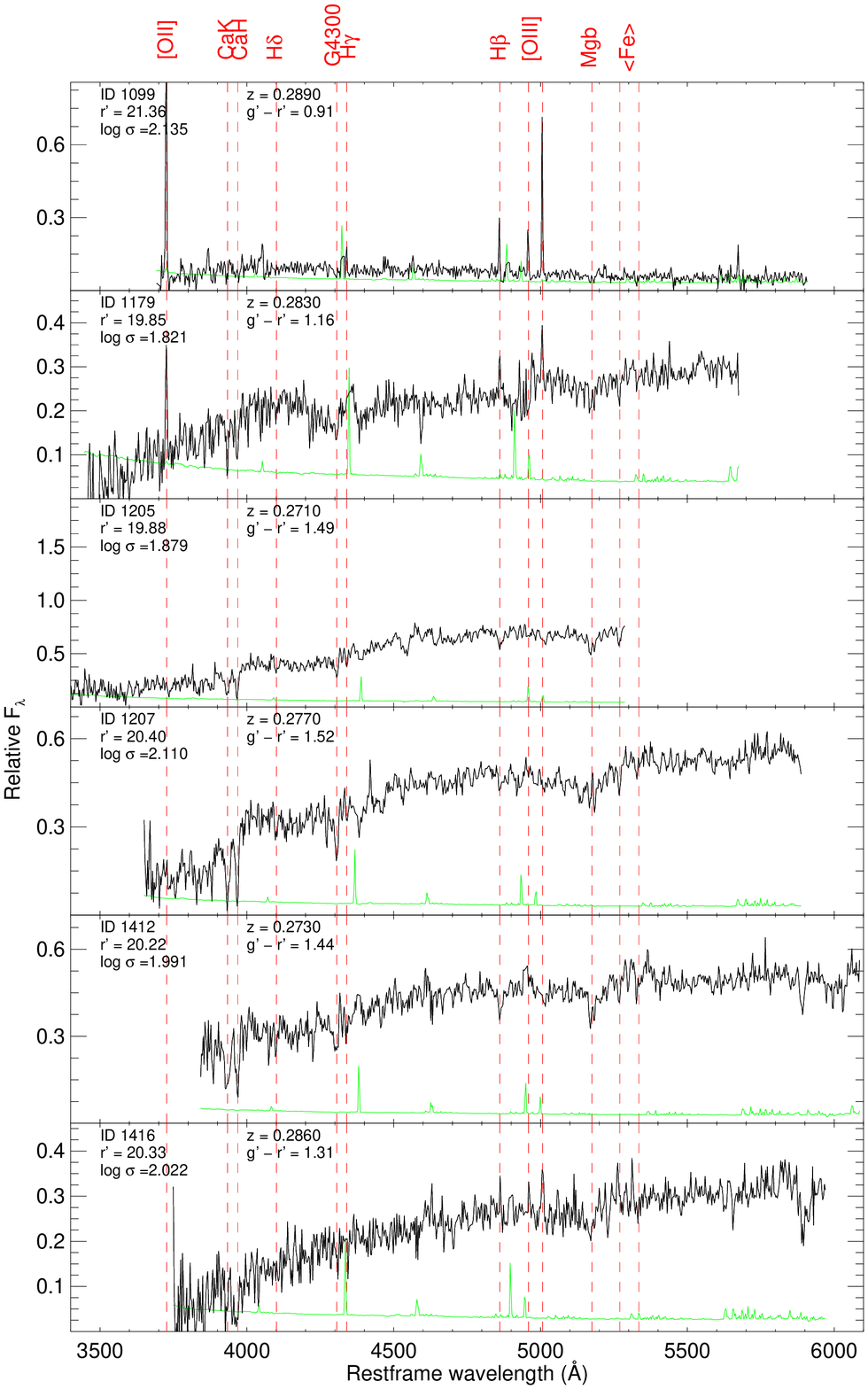}
    \figcaption{Continued.}

  \end{center}

\end{figure}

\begin{figure}

  \begin{center}

    \plotone{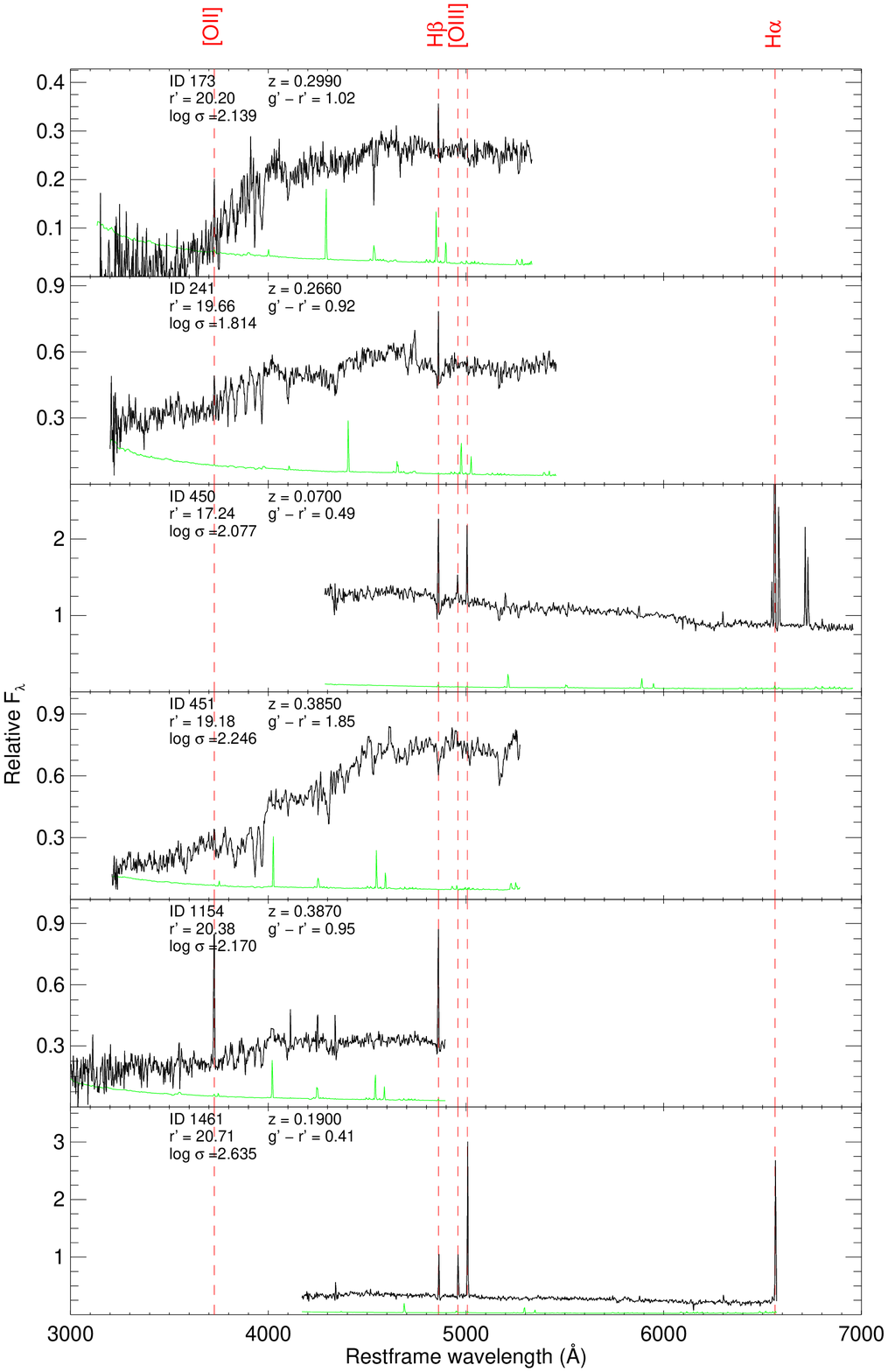}
    \figcaption{Extracted 1D spectra of non-members in the rest
    frame. The black line represents the flux-calibrated spectrum, the
    green line is noise multiplied by a factor of two for
    clarity. The positions of the most prominent emission lines are
    labeled with vertical dashed lines.\label{fig:sp2}}

  \end{center}

\end{figure}

\setcounter{figure}{10}
\begin{figure}

  \begin{center}

    \plotone{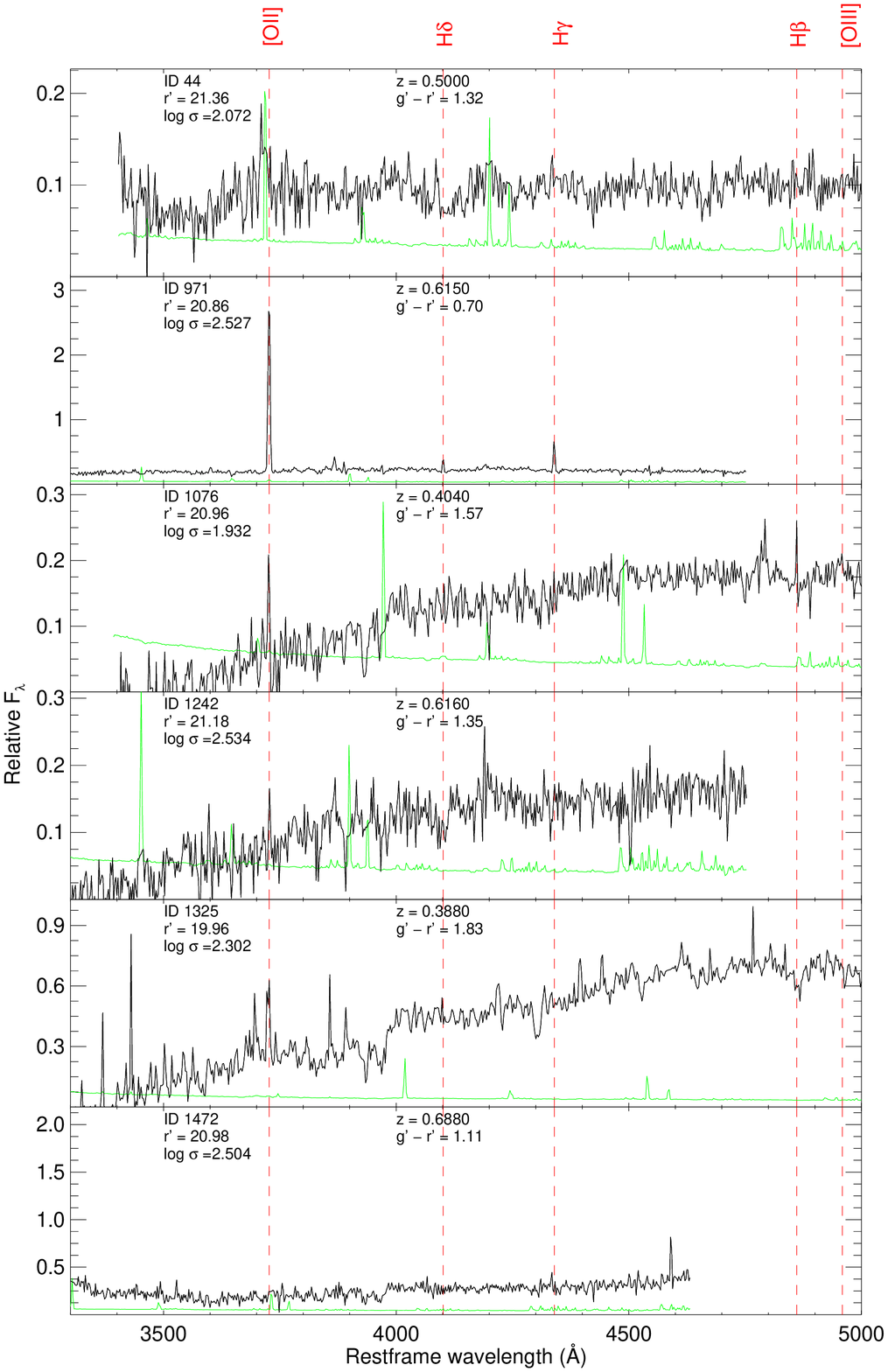}
    \figcaption{Continued.}

  \end{center}

\end{figure}

\begin{figure}

  \begin{center}

    \epsscale{1.2}
    \plotone{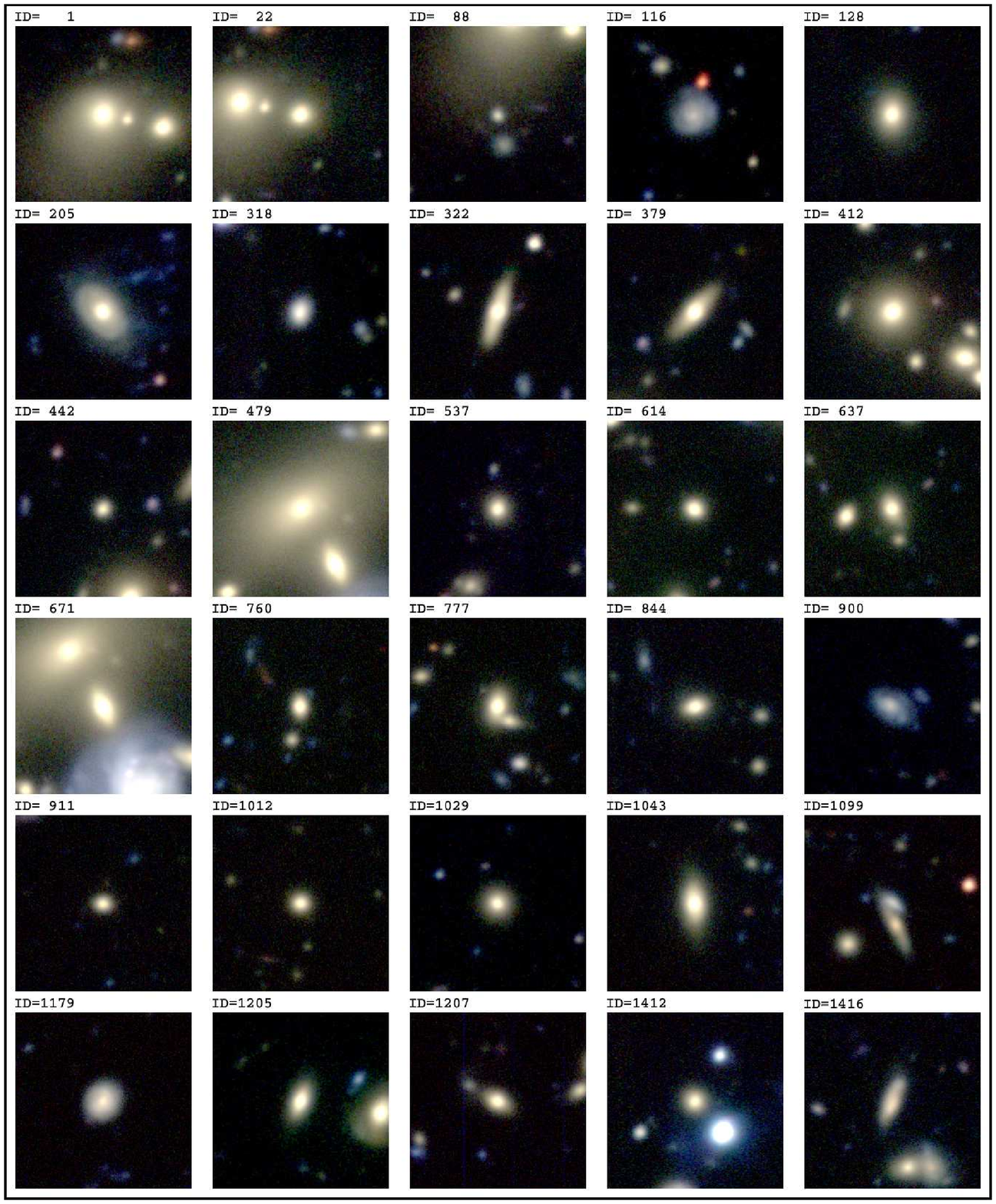}

    \figcaption{Postage stamps of cluster members in \rxj. Each frame
    is \arcsd{17}{5} by \arcsd{17}{5}, corresponding to $75
    \mathrm{kpc} \times 75 \mathrm{kpc}$ at $z=0.28$, with the
    spectroscopic galaxy at the centre. North is up, East is left. The
    stamps are produced from the GMOS-N \gfil, \rfil, \ifil \
    images. The galaxy ID is indicated at the top left of each
    stamp.\label{fig:crm}}

   \end{center}

\end{figure}

\begin{figure}

  \begin{center}

    \epsscale{1.2}
    \plotone{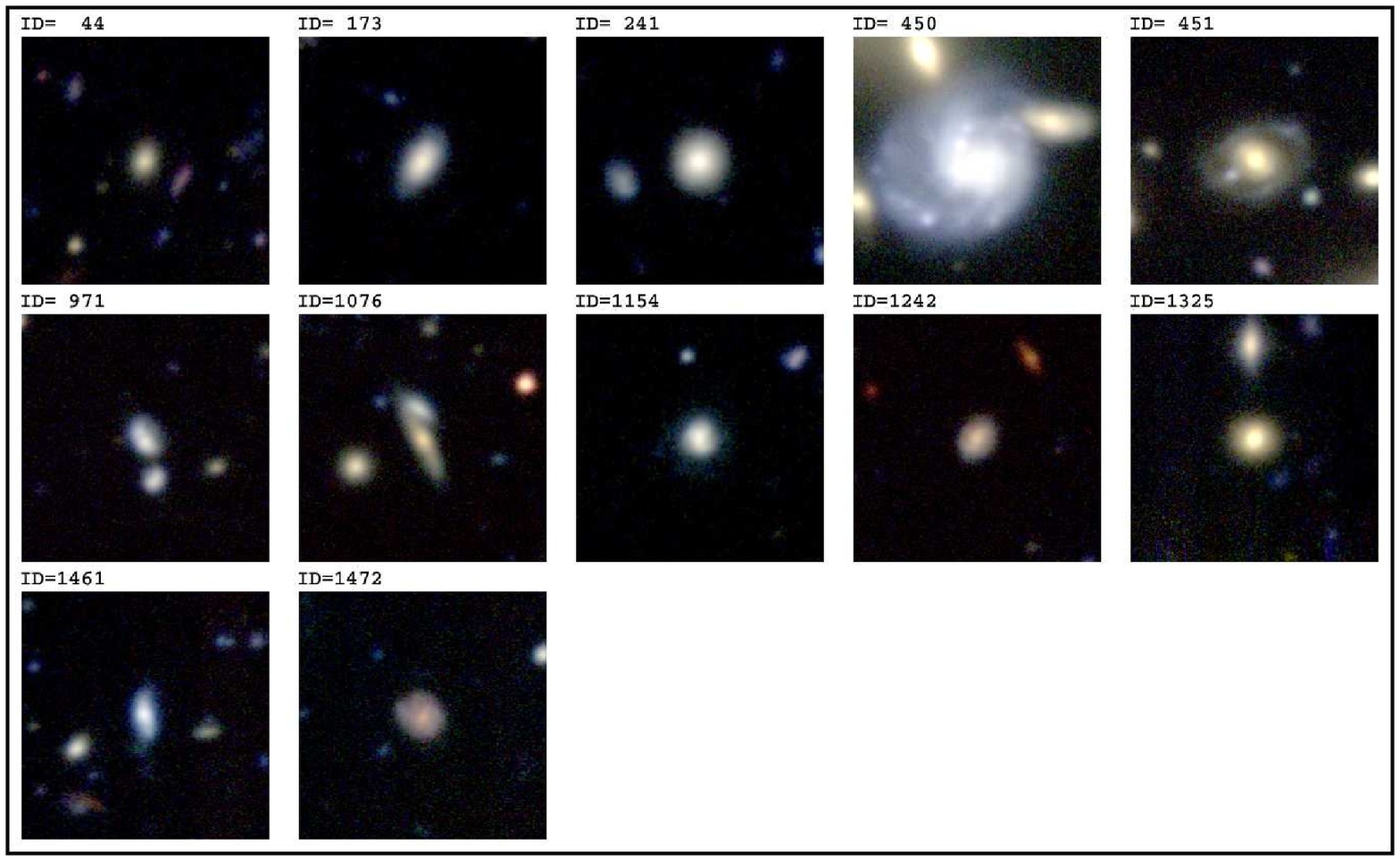}

    \figcaption{Color images of the non-cluster members in the
    spectroscopic sample. The images are equivalent to those in
    Figure~\ref{fig:crm}.\label{fig:crn}}

   \end{center}

\end{figure}

\typeout{LaTeX Warning: Label(s) may have changed. Rerun} 

\end{document}